\documentclass[twocolumn,aps,prx,superscriptaddress]{revtex4-2}
\usepackage[T1]{fontenc}
\usepackage[latin9]{inputenc}
\usepackage{textcomp}
\usepackage{bbding}
\usepackage{amsmath}
\usepackage{amssymb}
\usepackage{graphicx}
\usepackage{esint}
\usepackage{xcolor}
\usepackage[pdftex,colorlinks,linkcolor=blue,citecolor=blue,urlcolor=blue,filecolor=blue,breaklinks=true]{hyperref}

\makeatletter
\@ifundefined{textcolor}{}
{%
 \definecolor{BLACK}{gray}{0}
 \definecolor{WHITE}{gray}{1}
 \definecolor{RED}{rgb}{1,0,0}
 \definecolor{GREEN}{rgb}{0,1,0}
 \definecolor{BLUE}{rgb}{0,0,1}
 \definecolor{CYAN}{cmyk}{1,0,0,0}
 \definecolor{MAGENTA}{cmyk}{0,1,0,0}
 \definecolor{YELLOW}{cmyk}{0,0,1,0}
}

\makeatother

\begin{document}
\title{Minimalistic and Scalable Quantum Reservoir Computing Enhanced with Feedback}
\author{Chuanzhou Zhu}
\email{chuanzhouzhu@arizona.edu}
\affiliation{Wyant College of Optical Sciences, University of Arizona, Tucson, Arizona, USA}
\author{Peter J. Ehlers}
\affiliation{Wyant College of Optical Sciences, University of Arizona, Tucson, Arizona, USA}
\author{Hendra I. Nurdin}
\affiliation{School of Electrical Engineering and Telecommunications, University of New South Wales, Sydney, Australia}
\author{Daniel Soh}
\email{danielsoh@arizona.edu}
\affiliation{Wyant College of Optical Sciences, University of Arizona, Tucson, Arizona, USA}


\begin{abstract}
\begin{description}

\item[Abstract] Quantum Reservoir Computing (QRC) leverages quantum systems to perform complex computational tasks with exceptional efficiency and reduced energy consumption. We introduce a minimalistic QRC framework utilizing as few as five atoms in a single-mode optical cavity, combined with continuous quantum measurement. The system is conveniently scalable, as newly added atoms naturally couple with existing ones via the shared cavity field. To achieve high computational expressivity with a minimal reservoir, we include two critical elements: reservoir feedback and polynomial regression. Reservoir feedback modifies the reservoir's dynamics without altering its internal quantum hardware, while polynomial regression nonlinearly enhances output resolution. We demonstrate significant QRC performance in memory retention and nonlinear data processing through two tasks: predicting chaotic time-series data via the Mackey-Glass task and classifying sine-square waveforms. This framework fulfills QRC's objectives to minimize hardware size and energy consumption, marking a significant advancement in integrating quantum physics with machine learning technology.

\end{description}
\end{abstract}

\maketitle

\section{Introduction}

Quantum reservoir computing (QRC) has emerged as a groundbreaking machine learning framework, capable of addressing intricate tasks with remarkable efficiency and minimal energy consumption \cite{Fujii2017,chen2019learning,CNY20,Dudas2023,Govia2021,Bravo2022,Hulser2023,Nokkala2021,Pena2021,Xia2022,Mujal2023,Fry2023,Kalfus2022,Yasuda23,Beni2023,Markovic2019,Lin2020,kornjaca2024}. Some of these tasks fall into well-established categories in machine learning, such as time series analysis \cite{Fujii2017,Dudas2023,Govia2021} and computer vision \cite{kornjaca2024}, providing a clear benchmark for comparing the performance of QRC against classical reservoir computing (CRC) \cite{Tanaka2019,Appeltant2011,Pathak2018,Angelatos2021,Chen2022,Gauthier2021}. In addition, various QRC approaches and its static variants that process vectors rather than time series data, called quantum random kitchen sinks or quantum extreme learning machines \cite{Innocenti23}, have been employed to tackle tasks in the field of quantum research, such as recognizing quantum entanglements \cite{Ghosh2019}, measuring dispersive currents \cite{Angelatos2021}, and predicting molecular structures \cite{Domingo2022}.

Proposals for implementing quantum reservoirs have been advanced using various platforms, including coupled networks of qubits \cite{Fujii2017,chen2019learning,CNY20,Domingo2022,Ghosh2021,Pena2021,Xia2022,Yasuda23,McMahon2024}, fermions \cite{Ghosh2019}, harmonic oscillators \cite{Nokkala2021}, Kerr nonlinear oscillators \cite{Angelatos2021}, Rydberg atoms \cite{Bravo2022}, microparticles \cite{Cichos2024}, chiral magnets \cite{Kurebayashi20231,Kurebayashi20232}, optical pulses \cite{Beni2023}, artificial spin systems \cite{Kurebayashi2022}, chemical reaction networks \cite{Baltussen2024}, and phonon and magnon modes \cite{Scherbakov2023}. Increasing the number of physical sites, such as qubits, oscillators, or atoms, within a quantum network can dramatically expand the Hilbert space by exponentially increasing the number of quantum basis states. Each of these basis states functions as a node within the quantum neural network, mirroring the role of a node in a classical neural network \cite{Fujii2017} or an optical neural network \cite{Wang2022,ma2023}. The exponential scaling of basis states plays a critical role in making the system's dynamics nonlinear and complex. To ensure QRC performance, tens of physical sites are typically necessary for constructing a quantum reservoir \cite{kornjaca2024}. However, a significant challenge in scaling up a quantum reservoir is the considerable difficulty of establishing connections between each newly added site and all existing sites within a quantum network. 

To extract information from quantum reservoirs, the observable readouts have been proposed through various methods, including measuring probabilities on quantum basis states \cite{Dudas2023}, excitations of qubits \cite{Fujii2017}, occupations on lattice sites \cite{Ghosh2019}, and energies of oscillators \cite{Angelatos2021}. However, these conventional measurement techniques often require quantum tomography, which necessitates the complete destruction of the quantum reservoir to obtain readouts at each time step. This process demands a substantial number of repeated time evolutions, incurring a considerable overhead. To overcome this time complexity issue, several online protocols for QRC have been proposed. These include: online measurement based on partial readout and subsequent resetting of readout qubits to preserve memory of past inputs \cite{CNY20,Yasuda23,Hu2024}; weak and projective measurements to extract information accurately without hindering memory \cite{Mujal2023}; a feedback-driven QRC framework that feeds measurement outcomes back into the reservoir to restore memory of prior inputs \cite{Kobayashi2024}; and the application of singular value decomposition and data-filtering techniques to optimize the training of the finite-sampled QRC framework \cite{Ahmed2025}.

In this paper, we present a minimalistic quantum reservoir made of up to five atoms and a single-mode cavity field, where the input is practically embedded in external coherent laser driving. The connections among atoms can be more conveniently constructed compared to prior proposals, since newly added atoms automatically couple to existing atoms through their mutual connections with the cavity field, ensuring the practical scalablilty of the system.

In small hardware systems, maximizing computational expressivity is imperative. To achieve this, we adopt two main strategies: adjustable feedback and polynomial regression. First, we introduce a feedback mechanism that feeds observable readouts back into the reservoir as input. This approach is related to the feedback formalism in classical echo state networks (ESNs) \cite{Peter2023} and biological systems \cite{Maass2007}. This mechanism provides an excellent solution: it allows us to modify the reservoir's overall dynamics and significantly enhance its computational expressivity without altering its internal hardware details. Second, we utilize polynomial combinations of readouts to enrich the complexity of the output, leading to another significant performance boost. The observable readouts in our scheme are practically obtained through a continuous quantum measurement, where the cavity field provides two readouts linked to two photonic quadratures, and each atom provides two readouts linked to two atomic spin channels \cite{Wiseman2010,BvHJ07,Wei2008,Ruskov2010,Hacohen2016,Ochoa2018,Fuchs2001}. By adopting a continuous measurement scheme that accounts for measurement back-action on the reservoir, we eliminate the need for the costly tomography measurements used in previous methods. Our proposed approach, implemented on minimalistic quantum hardware, combines a feedback mechanism, polynomial regression, and continuous measurement. Together, these elements aim to achieve highly efficient QRC with minimal energy consumption.
 
We evaluate the QRC performance on two time-series tasks: the Mackey-Glass forecasting task \cite{Fujii2017}, which demands long-term memory for predicting the future trend of an input function while providing the fading memory property simultaneously, and the sine-square waveform classification task \cite{Dudas2023,Markovic2019}, which requires nonlinearity of the reservoir to capture sudden, high-frequency shifts in a linearly inseparable input dataset. The significant performance enhancement associated with the increased number of atoms, the feedback mechanism, and the polynomial regression are demonstrated by the diminishing gaps between the actual and target outputs in the two tasks.

\section{Results}

\begin{figure*}
\includegraphics[width=1.0\linewidth]{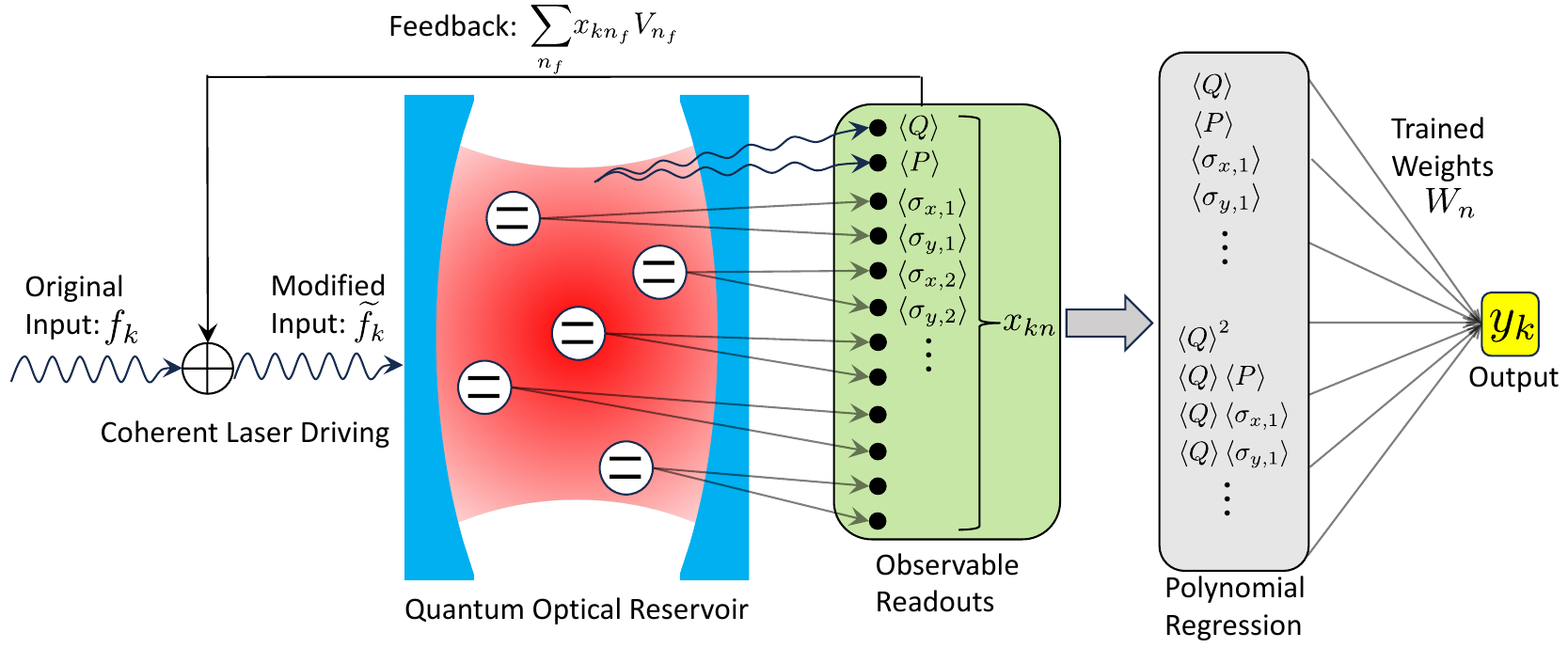}
\caption{Setup of quantum optical reservoir computing with feedback. The input function is coherently integrated into the driving field of an optical cavity. The reservoir consists of atoms inside the cavity, exhibiting diverse detunings and coupling strengths across various spatial positions. Readouts are obtained via continuous measurements of photonic quadratures and atomic spin channels, denoted by $x_{kn}$ with $n$ the index of the readout channel.
The feedback mechanism is implemented by modifying the input using several readouts, $x_{kn_{f}}$, where $n_{f}$ indexes the readouts used for feedback. The parameter $V_{n_{f}}$ controls the strength of the $n_{f}$-th feedback channel. 
A polynomial regression is then applied to map the readouts to the output with weights $W_{n}$. The parameters $V_{n_{f}}$ and $W_{n}$ are trained through a machine learning process to enhance the performance of QRC.}
\label{Scheme}
\end{figure*}

\subsection{Quantum Reservoir Computing Setup}

We consider a QRC framework with a feedback mechanism, as illustrated in Fig.~\ref{Scheme}. In this setup, input samples are encoded in a real-valued time series $f_{k}$, where the integer $k$ denotes the discrete time index. The input is processed by the quantum reservoir, and the corresponding readouts $x_{kn}$ are acquired through continuous quantum measurement, with $n$ representing the index of the readout channel. The feedback mechanism is implemented by using a subset of the readouts, denoted as $x_{kn_{f}}$, to modify the input as 
\begin{equation}
\widetilde{f}_{k}=f_{k}+\underset{n_{f}}{\sum}x_{kn_{f}}V_{n_{f}},\label{eq:ftilde}
\end{equation}
where $n_{f}$ indexes the feedback channels. The readouts are then linked to the output to perform a time-series forecasting or classification task via polynomial regression, where the associated weights are denoted as $W_{n}$. The machine-learning parameters $V_{n_{f}}$ and $W_{n}$ are trained using global optimization and pseudoinverse regression, respectively, to optimize the QRC performance. Note that the framework of Eq.~(\ref{eq:ftilde}) represents a general feedback formalism applicable to any quantum reservoir in which the input is modified by its own measurement outcomes. In the quantum optical reservoir considered below, $\widetilde{f}_{k}$ is implemented via coherent driving of a cavity, while $x_{kn_{f}}$ corresponds to continuous measurement of observable expectation values.

We employ a quantum optical system encompassing $N_{atom}$ two-level atoms inside a single-mode optical cavity \cite{Jaynes1963,Tavis1968} as the quantum reservoir. The time-independent part of the Hamiltonian is expressed as 
\begin{equation}
H_{0}=\omega_{c}c^{\dagger}c+\underset{i}{\sum}\omega_{i}\sigma_{i}^{\dagger}\sigma_{i}+\underset{i}{\sum}g_{i}\left(c^{\dagger}\sigma_{i}+c\sigma_{i}^{\dagger}\right),\label{eq:H0}
\end{equation}
where $c$ represents the photon annihilation operator, $\sigma_{i}=\left|g\right\rangle \left\langle e\right|_{i}$ denotes the lowering operator of the $i$-th atom with $\left|g\right\rangle $ ($\left|e\right\rangle $) representing the ground (excited) state, $\omega_{c}$ ($\omega_{i}$) describes the detuning between the coherent driving and the cavity (atomic) frequencies, and $g_{i}$ is the electric-dipole coupling strength between the $i$-th atom and the cavity mode. For QRC, it is essential to select either various detuning $\omega_{i}$ or various coupling strength $g_{i}$ to prompt the atoms to produce non-identical memories, thus enhancing their overall capability. This model can be practically implemented using cold atoms in an optical cavity \cite{Brennecke2007,Niemczyk2010}, where optical tweezers can be used to trap and measure individual atoms at different positions \cite{Kaufman2021,Ye2023}. Alternatively, quantum dots can also be utilized to induce random positioning, detuning, and coupling \cite{Schmidt2007}. 

The feedback-modified input $\widetilde{f}_{k}$ is encoded into the time-dependent coherent laser driving as
\begin{equation}
H_{1}\left(t_{k}\leqslant t<t_{k+1}\right)=i\epsilon\widetilde{f}_{k}\left(c-c^{\dagger}\right),\label{eq:H1}
\end{equation}
where $\epsilon$ denotes the driving strength. Time is discretized as $t_{k}=k\Delta t$, with $\Delta t$ representing the time step. During the time interval from $t_{k}$ to $t_{k+1}$, the input driving to the quantum reservoir remains constant. 

The readouts $x_{kn}$ are determined through continuous measurement of quantum observables. The observables of the cavity field stem from the homodyne detection of two orthogonal quadratures \cite{Wiseman2010,BvHJ07,Nurdin14}
\begin{align}
Q & =c+c^{\dagger},\label{eq:Q}\\
P & =i\left(c-c^{\dagger}\right),\label{eq:P}
\end{align}
and the observables of the atomic spontaneous emission are associated with the Pauli operators \cite{Wiseman2001}
\begin{align}
\sigma_{x,i} & =\sigma_{i}+\sigma_{i}^{\dagger},\label{eq:sigmax}\\
\sigma_{y,i} & =i\left(\sigma_{i}-\sigma_{i}^{\dagger}\right).\label{eq:sigmay}
\end{align}
It has been demonstrated that these observables of the cavity and atoms can be simultaneously measured \cite{Wei2008,Ruskov2010,Hacohen2016,Ochoa2018}. By averaging over a large number of measurement trajectories, the readouts are correlated with the expectation values as $x_{k1}=\left\langle Q\right\rangle _{k}$, $x_{k2}=\left\langle P\right\rangle _{k}$, $x_{k3}=\left\langle \sigma_{x,1}\right\rangle _{k}$, $x_{k4}=\left\langle \sigma_{y,1}\right\rangle _{k}$, and so forth. This leads to $N_{readouts}=2N_{atom}+2$, where $N_{readouts}$ is the number of available readout channels and $N_{atom}$ is the number of atoms. The expectation values are calculated as $\left\langle Q\right\rangle _{k}={\rm Tr}\left[\rho\left(t_{k}\right)Q\right]$, for instance, where $\rho\left(t_{k}\right)$ represents the density operator at time $t_{k}$, which evolves according to the dynamics of the reservoir simulated by the quantum master equation. The dynamics of the quantum reservoir with continuous quantum measurement are elaborated in the ``Methods'' section. 

To implement the feedback mechanism, a subset of readouts, denoted as $x_{kn_{f}}$, is selected to serve as feedback. These readouts are combined with the original input $f_{k}$ to generate the modified input $\widetilde{f}_{k}$, as described in Eq.~(\ref{eq:ftilde}). At time $t_{k}$, the modified input enters the quantum reservoir in terms of the coherent driving shown in Eq.~(\ref{eq:H1}). The evolution of the quantum reservoir from $t_{k}$ to $t_{k+1}$, simulated by the master equation, produces the readouts $x_{k+1,n_{f}}$ at time $t_{k+1}$, which are then used to modify the input $\widetilde{f}_{k+1}$ for the next feedback cycle. This reservoir feedback scheme can be viewed as a \textit{partial} hidden state of a recurrent-structured reservoir using a fanned-out readout. However, compared to those in classical recurrent structures, our readouts, $x_{kn_{f}}$, are generated by a nonlinear quantum system with a large number of hidden basis states. This ensures significant QRC performance even with a small number of readouts since our simple feedback captures the complex dynamics occurring in a high-dimensional Hilbert space. Moreover, it is only partial feedback since not all of the expectation values of basis operators that span the space of all observables of the quantum system are fedback. 

In the ``Methods'' section, we present an online protocol for the feedback mechanism based on continuous quantum measurement. This approach eliminates the need for multiple copies of the quantum system and repeated measurements to estimate the expectation values of observables at each time step. By enabling multiple continuous measurement trajectories on a single quantum system, the protocol avoids the use of destructive measurements.

After completing all feedback cycles, polynomial regression is employed by incorporating both the linear and quadratic terms of the observable expectation values into the readouts, resulting in $N_{readouts}=2N_{atom}^{2}+7N_{atom}+5$. The terms encompassed in polynomial regression are exemplified in the grey box in Fig.~\ref{Scheme}. The mapping from the readouts to the output is given by
\begin{equation}
y_{k}=\underset{n}{\sum}x_{kn}W_{n},\label{eq:yk}
\end{equation}
where $W_{n}$ are the regression coefficients to be trained. In regular linear regression, the regressors $x_{kn}$ include only the linear terms of the expectation-value measurements, while in polynomial regression, they include both the linear and quadratic terms. 

The goal of training is to optimize the parameters $V_{n_{f}}$ and $W_{n}$ to minimize the normalized root mean square error (NRMSE) defined by
\begin{equation}
{\rm NRMSE}=\frac{1}{\bar{y}_{max}-\bar{y}_{min}}\sqrt{\frac{\underset{k}{\sum}\left(y_{k}-\bar{y}_{k}\right)^{2}}{L}},\label{eq:NRMSE}
\end{equation}
where $L$ is the number of time steps in the training period and $\bar{y}_{k}$ represents the target output that captures some key features of the input. We adopt various global optimization methods to train $V_{n_{f}}$ and the Moore-Penrose pseudoinverse method to train $W_{n}$. The specific training methods are detailed in the ``Methods'' section. During the testing phase, the trained weights $V_{n_{f}}$ and $W_{n}$ are treated as fixed, and the NRMSE is computed to evaluate the QRC performance. 

The scale of QRC is determined by four factors: (i) the number of basis states spanning the Hilbert space, which scales as $N_{c}2^{N_{atom}}$, where $N_{c}$ is the number of involved photon Fock states and $N_{atom}$ is the total number of ``measured'' and ``unmeasured'' atoms; (ii) the number of readouts, analogous to the number of neurons in CRC, which is determined by the number of ``measured'' atoms and related to the index $n$, with a maximum availability of $2N_{atom}+2$; (iii) the number of feedbacks, which is decided by the number of readouts that are used to modify the input through the feedback mechanism, and associated with the index $n_{f}$; and (iv) the implementation of polynomial regression, which incorporates both linear and quadratic terms of the readouts.

\begin{figure}
\includegraphics[width=1.0\linewidth]{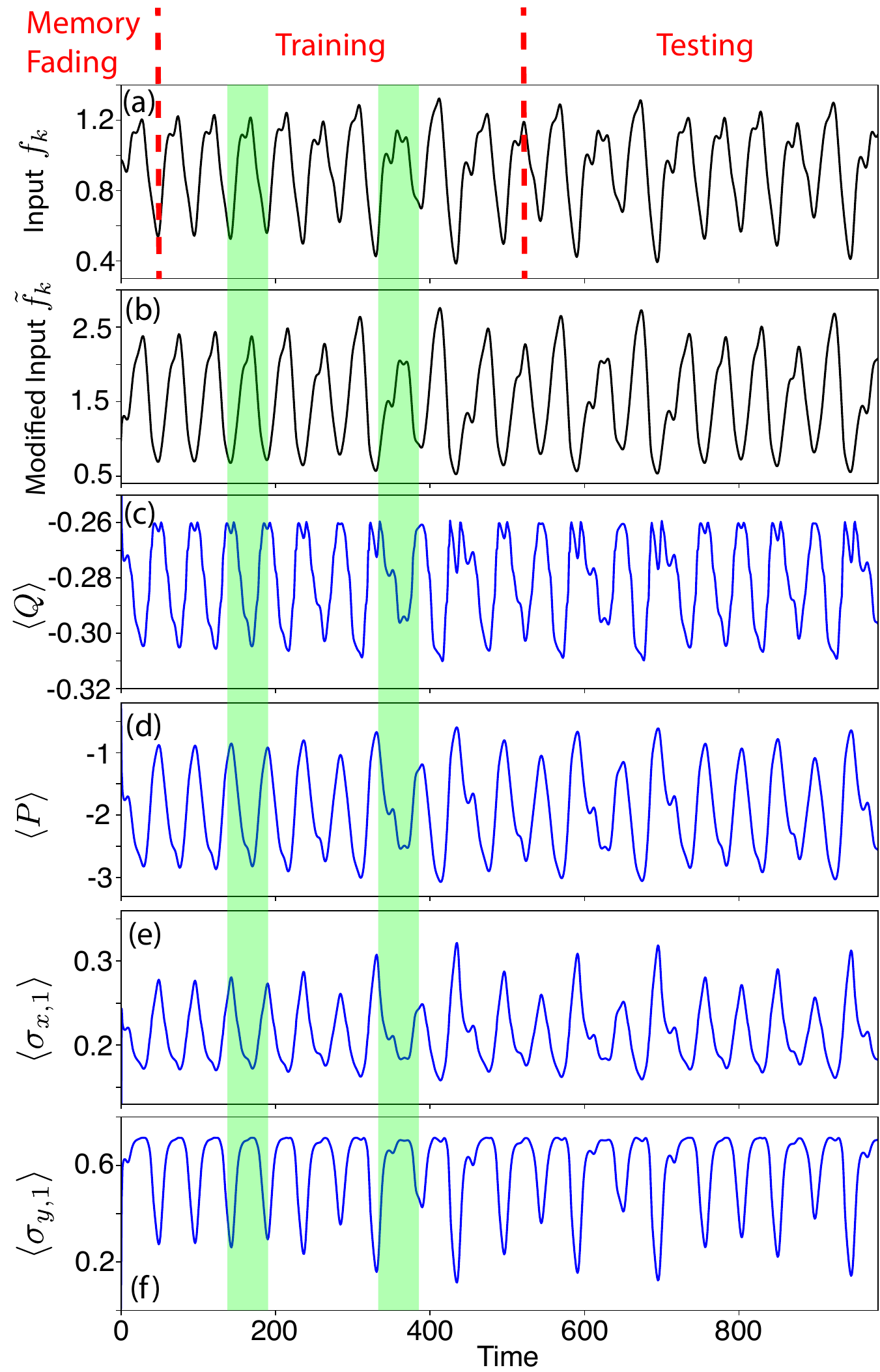}
\caption{Input and readouts for the Mackey-Glass task utilizing a single-atom reservoir with $\omega_{1}=20$ and $g_{1}=30$. 
(a) The input function to be processed, $f_{k}$, divided into memory fading, training, and testing regions. 
(b) The actual input applied to the reservoir, $\widetilde{f}_{k}$, modified by the feedback mechanism using all the 4 available readouts as feedback channels.
(c)-(f) The corresponding 4 readouts from photonic quadratures and atomic spin channels. 
The green shaded regions highlight the distinct responses of the reservoir to different waveforms in the input signal.  
Parameters: $Delay=20$, $\kappa=10$, $\omega_{c}=40$, and $\epsilon=20$.}
\label{Mackey_Glass_Input_Readouts}
\end{figure}

\subsection{Mackey-Glass Task}

The Mackey-Glass task serves as a test for long-term memory, demanding the reservoir to retain past information from the input function to forecast its future behavior. The chaotic nature of the Mackey-Glass time series makes it a challenging problem for machine learning models, forcing them to effectively learn the underlying patterns. The input is generated from the Mackey-Glass equation
\begin{equation}
\frac{df\left(t\right)}{dt}=\frac{\beta f\left(t-\tau_{M}\right)}{1+f^{10}\left(t-\tau_{M}\right)}-\gamma f\left(t\right),\label{eq:MGEq}
\end{equation}
where the parameters $\beta=0.2$, $\gamma=0.1$, and $\tau_{M}=17$ are commonly accepted standard values in the chaotic regime \cite{Fujii2017,Dudas2023,Hulser2023}. Following the convention used in Ref.~\cite{Hulser2023}, we apply a buffer by discarding the first $1000$ time units in Eq.~(\ref{eq:MGEq}). Thus, we consider the discretized function $f_{k}=f\left(t+1000\right)$ as the original input, with discretized time $t=k \Delta t$, as depicted in Fig.~\ref{Mackey_Glass_Input_Readouts}(a). The time series is sampled with a time step of $\Delta t=1$. The input $f_{k}$ is modified to $\widetilde{f}_{k}$ by the feedback system in each time step, as described by Eq.~(\ref{eq:ftilde}). The target output, $\bar{y}_{k}$, predicting the future of the input function with a time $Delay$, is constructed as $\bar{y}_{k}=f_{k+Delay}$. The readouts of observables illustrated in Fig.~\ref{Mackey_Glass_Input_Readouts}(c)-(f) show that the reservoir exhibits discernible responses to various input waveforms, which are exemplified by the green shadows corresponding to two different input waveforms. The time series is divided into three intervals for the purposes of memory fading, training and testing. The memory fading phase guarantees that during training and testing, the readouts depend solely on the input function rather than the initial state of the master equation. 

\begin{figure*}
\includegraphics[width=1.0\linewidth]{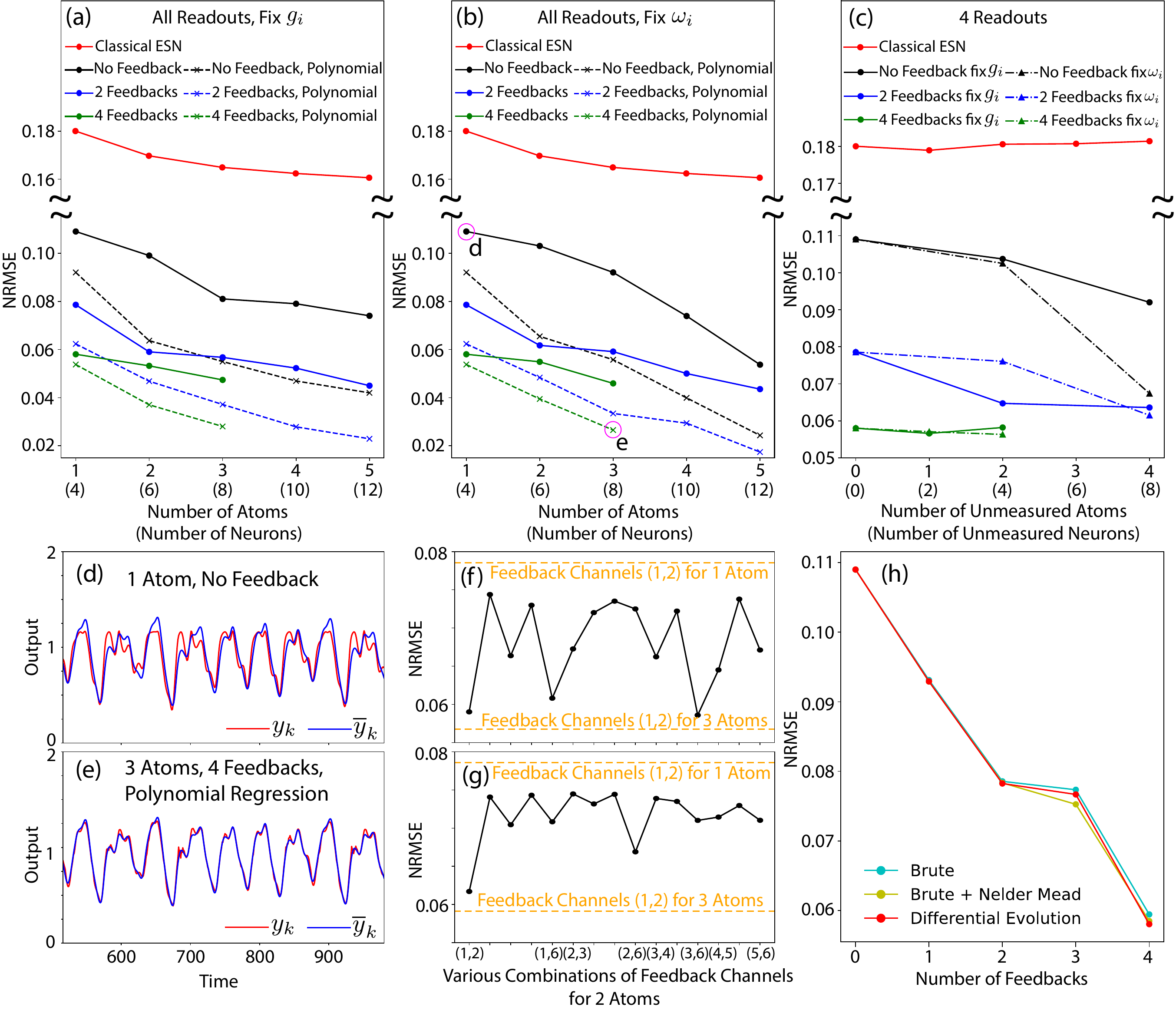}
\caption{Performance testing for the Mackey-Glass task.
(a)(b) The tested NRMSE plotted against the number of atoms (or neurons), using all available readouts from the cavity field and all atoms, with no feedback (black), $2$ feedbacks via $x_{k1}$ and $x_{k2}$ channels (blue), or $4$ feedbacks via $x_{k1}$ to $x_{k4}$ channels (green), where regular linear regression (solid lines) or polynomial regression (dashed lines) is applied. In panel (a), the coupling strength $g_i = 30$ is fixed for all atoms, and the detunings $\omega_i$ vary: $\omega_i = 20$ for one atom, $\omega_i = [0, 40]$ for two atoms, $\omega_i = [0, 20, 40]$ for three atoms, $\omega_i = [0, 10, 30, 40]$ for four atoms, and $\omega_i = [0, 10, 20, 30, 40]$ for five atoms. In panel (b), $\omega_i = 20$ is fixed for all atoms, and $g_i$ varies: $g_i = 30$ for one atom, $g_i = [10, 50]$ for two atoms, $g_i = [10, 30, 50]$ for three atoms, $g_i = [10, 20, 40, 50]$ for four atoms, and $g_i = [10, 20, 30, 40, 50]$ for five atoms. In panels (a)(b), classical reservoir computing (CRC) (red solid line) uses all available readouts from all neurons in classical echo state networks (ESN) (detailed in the ``Supplementary Information'').
(c) The tested NRMSE as a function of the number of unmeasured atoms (for QRC) or unmeasured neurons (for CRC), using only the $4$ readout channels (taking 4 measurements) from the cavity field and a particular atom with $g=30$ and $\omega=20$, where the total number of atoms is increased by fixing either $g_{i}$ (solid lines) or $\omega_{i}$ (dashed-dotted lines).
(d)(e) Actual (red) and target (blue) outputs from QRC utilizing one atom with no feedback with linear regression, and three atoms with $4$ feedbacks with polynomial regression, corresponding to the points marked by letters ``d'' and ``e'' in panel (b), respectively. 
(f)(g) The influence of different selections for the $2$ feedback channels out of the $6$ readout channels in two-atom QRC (black lines), where ($n$,$m$) denotes the feedback channels $x_{kn}$ and $x_{km}$ with $n<m$. The results from one-atom and three-atom QRC using feedback channels $x_{k1}$ and $x_{k2}$ (orange lines) are plotted for reference. Panel (f) fixes $g_i$ for all atoms, while panel (g) fixes $\omega_i$. 
(h) Comparison of three methods for training feedback parameters $V_{n_{f}}$: differential evolution, brute force, and brute force plus Nelder Mead, under varying numbers of feedbacks for one-atom QRC with $g_{1}=30$ and $\omega_{1}=20$ using linear regression.
Parameters: $Delay=20$, $\kappa=10$, $\omega_{c}=40$, $\epsilon=20$.}
\label{Mackey_Glass_Change_AtomNum}
\end{figure*}

The performance enhancement on account of the scalability of QRC is illustrated in Fig.~\ref{Mackey_Glass_Change_AtomNum}. Increasing the number of atoms boosts the performance by expanding both the size of the Hilbert space and the number of available readouts. Figures~\ref{Mackey_Glass_Change_AtomNum}(a)(b) show the results when the number of atoms is increased from $1$ to $5$, utilizing all available readout channels from the cavity field and all the atoms, with $g_i$ and $\omega_i$ fixed, respectively. This leads to a corresponding increase in the number of readouts (neurons) from $4$ to $12$. The remarkable performance enhancement associated with the feedback mechanism is depicted by the blue ($2$ feedbacks) and green ($4$ feedbacks) lines in Figs.~\ref{Mackey_Glass_Change_AtomNum}(a)(b), while the results from polynomial regression are shown by the dashed lines. A comparison between two extreme cases, one-atom QRC without feedback finished by linear regression and three-atom QRC with 4 feedbacks finished by polynomial regression, is illustrated in Figs.~\ref{Mackey_Glass_Change_AtomNum}(d)(e). This comparison demonstrates a substantial performance improvement resulting from the combination of all the scalable factors. 

For CRC based on classical echo state networks (ESN), performance enhancement resulting from the increased number of neurons is plotted by the red solid lines in Figs.~\ref{Mackey_Glass_Change_AtomNum}(a)(b), with details discussed in the ``Supplementary Information''. The primary advantage of QRC over CRC lies in the entanglement of atomic states through coupling with a shared cavity field. The exponentially increased number of atomic basis states achieves a high-dimensional Hilbert space for QRC, despite only a relatively small number of observable readouts are being measured.

The effect of solely increasing the dimension of the QRC Hilbert space is demonstrated in Fig.~\ref{Mackey_Glass_Change_AtomNum}(c), where the number of measured readouts is fixed at $4$ (including $2$ from the cavity and $2$ from an atom), while additional unmeasured atoms are added for the only purpose of increasing the number of basis states. Performance improvements are observed in both the no-feedback and 2-feedback cases, indicating the inherent connections and collaborations between the measured atom and the added atoms. The effect tends to saturate in the 4-feedback case as the use of all the $4$ available feedback channels tends to dominate the performance. The similar effect has been observed for Ising models of QRC \cite{Fujii2017,chen2019learning}. A reason for this effect could be that, for the particular tasks considered, more complex fading memory maps generated by the QRC as the Hilbert space is increased are able to better capture features in the task to be learned. However, the improvement is expected to eventually plateau for a high enough dimension of the Hilbert space. In Fig.~\ref{Mackey_Glass_Change_AtomNum}(c), we also compare the results with the classical ESN. One of the most striking features of QRC over CRC is that increasing the number of unmeasured neurons improves the performance in QRC (black and blue solid lines in Fig.~\ref{Mackey_Glass_Change_AtomNum}(c)), whereas the CRC does not improve the performance even if we increase the number of unmeasured neurons (flat red line in Fig.~\ref{Mackey_Glass_Change_AtomNum}(c)). In the ``Supplementary Information'', the performance of CRC is evaluated by fixing the measured neurons while increasing the number of unmeasured neurons, and no performance improvement is observed for CRC.

The QRC performance can also be influenced by the selection of feedback channels, due to the different origins of the readouts from the cavity and atoms with distinct parameters. In the two-atom, 2-feedback cases (corresponding to the second blue dots on the blue solid lines in Figs.~\ref{Mackey_Glass_Change_AtomNum}(a)(b)), the first $2$ readouts, $x_{k1}=\left\langle Q\right\rangle $ and $x_{k2}=\left\langle P\right\rangle $, are chosen as feedback channels. However, there are a total of $6$ possible feedback options, which also include $x_{k3}=\left\langle \sigma_{x,1}\right\rangle $, $x_{k4}=\left\langle \sigma_{y,1}\right\rangle $, $x_{k5}=\left\langle \sigma_{x,2}\right\rangle $, and $x_{k6}=\left\langle \sigma_{y,2}\right\rangle $. Figures.~\ref{Mackey_Glass_Change_AtomNum}(f)(g) demonstrate the influence of the different selections for the $2$ feedbacks, with $g_{i}$ and $\omega_{i}$ fixed, respectively. A comparison with the one-atom and three-atom results shows that the choice of feedback channels has a smaller effect on performance than the number of atoms, confirming the generality of the result shown in Figs.~\ref{Mackey_Glass_Change_AtomNum}(a)(b).

The feedback parameters $V_{n_{f}}$ are optimized using three global optimization methods: differential evolution, brute force, and brute force combined with the Nelder-Mead algorithm. The corresponding test results are shown in Fig.~\ref{Mackey_Glass_Change_AtomNum}(h), with further details provided in the ``Methods'' section. Our results suggest that the feedback mechanism yields a noticeable improvement in QRC performance, even when only a small number of $V_{n_{f}}$ values are explored within a constrained parameter space and only a single feedback channel is activated. Thus, rigorous global optimization of $V_{n_{f}}$ is not necessary to achieve substantial performance enhancement. For example, in the brute force method, the $V_{n_{f}}$ space is discretized with a step size of $0.5$ and the search boundaries are limited to $V_{n_{f}}\in\left[-3,3\right]$. Referring to these settings and the 1-feedback case shown in Fig.~\ref{Mackey_Glass_Change_AtomNum}(h), the NRMSE improves from approximately $0.11$ to $0.09$ by evaluating only $13$ different values of $V_{n_{f}}$. The NRMSE profile as a function of a single feedback parameter $V_{m}$--with all other feedback channels deactivated (i.e., $V_{n_{f} \ne m} = 0$)--is presented in Fig.~S1(a-d) of the ``Supplementary Information.'' These plots demonstrate performance enhancement resulted from various options for $V_{m}$. Additionally, as shown in Fig.~S1, high-frequency oscillations in the readouts, related to a limit-cycle-like behavior, are observed in certain parameter regions.

\begin{figure}
\includegraphics[width=1.0\linewidth]{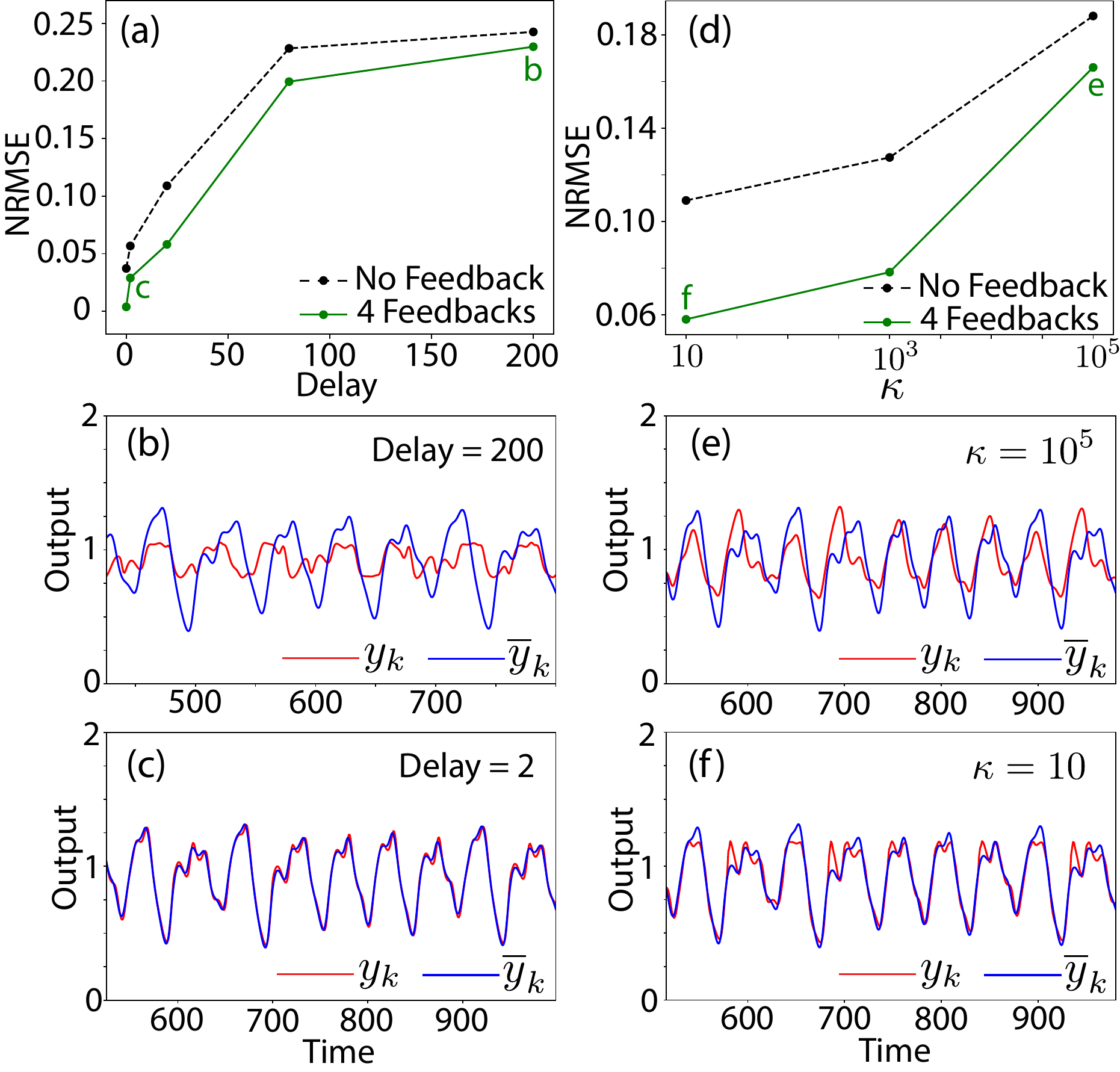}
\caption{Performance testing for the Mackey-Glass task using one-atom QRC with various $Delay$ and decay rates $\kappa$.
(a) NRMSE as a function of $Delay$, with no feedback or 4 feedbacks, for $\kappa=10$.
(b)(c) The actual (red) and target (blue) outputs for $Delay=200$ and $Delay=2$ with $\kappa=10$, corresponding to the points marked by letters ``b'' and ``c'' in panel (a), respectively.
(d) NRMSE as a function of $\kappa$, with no feedback or 4 feedbacks, for $Delay=20$.
(e)(f) The actual (red) and target (blue) outputs for $\kappa=10^{5}$ and $\kappa=10$ with $Delay=20$, corresponding to the points marked by letters ``e'' and ``f'' in panel (d), respectively.
Parameters: $\omega_{1}=20$, $g_{1}=30$, $\omega_{c}=40$, $\epsilon=20$.}
\label{Mackey_Glass_Change_Delay_Kappa}
\end{figure}

The impacts of $Delay$ and the decay rate $\kappa$ are shown in Fig.~\ref{Mackey_Glass_Change_Delay_Kappa}. In the Mackey-Glass task, a longer $Delay$ necessitates a stronger memory for the reservoir to ``remember'' more past information from the input in order to forecast the future output, making the task increasingly challenging. The decay rate $\kappa$ characterizes the loss of quantum memory caused by quantum dissipation, as define in the stochastic master equation introduced in the ``Methods'' section. A larger $\kappa$ tends to cause the reservoir to ``forget'' input information more quickly due to photon leakage and atomic spontaneous emission. However, a larger $\kappa$ is beneficial for stronger measurements with reduced uncertainties in readouts. Therefore, selecting a balanced $\kappa$ could be crucial in practice.

\begin{figure}
\includegraphics[width=1.0\linewidth]{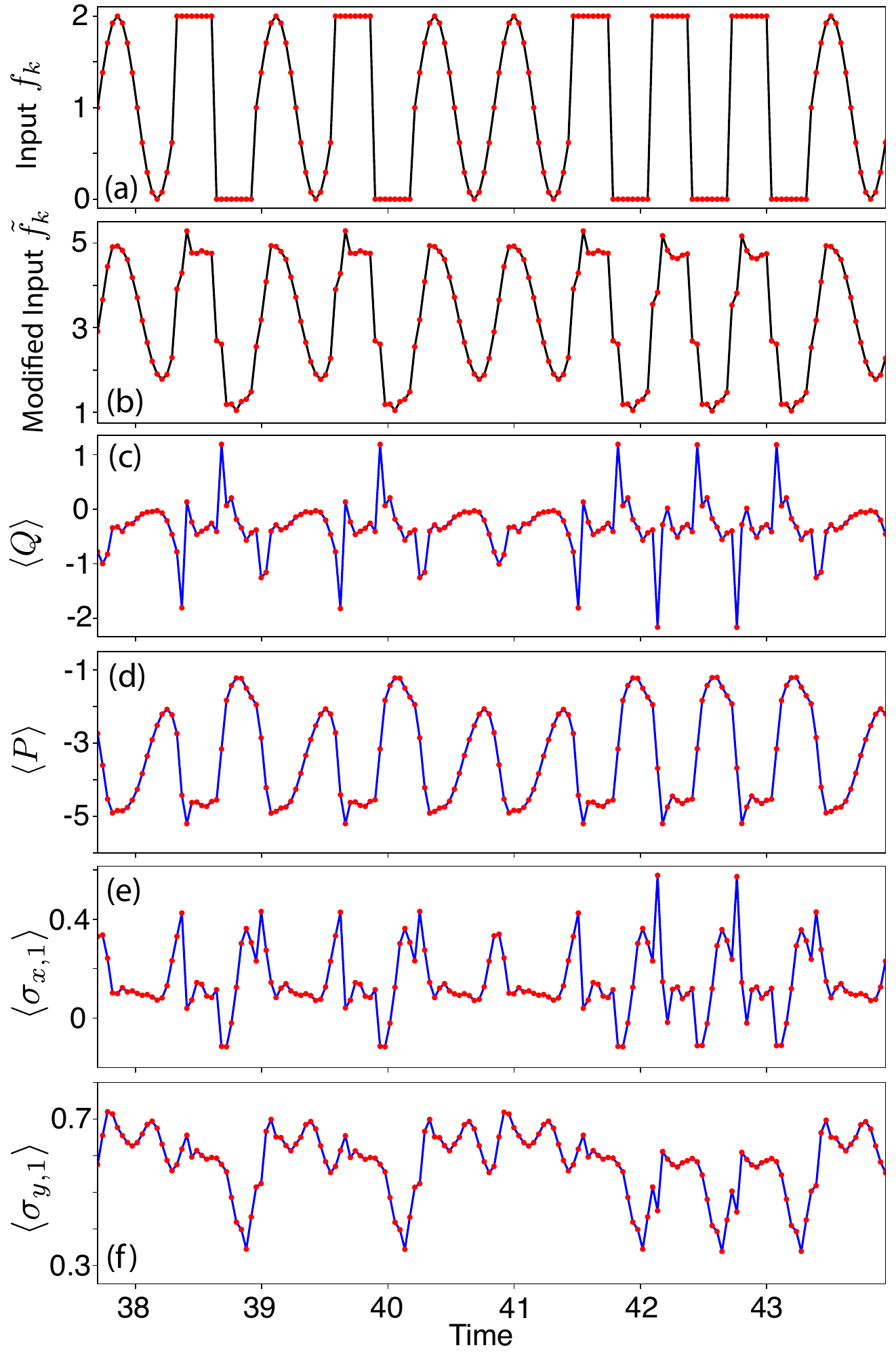}
\caption{Input and readouts for the sine-square waveform classification task. A total of $110$ random waveforms are sent as input, with $10$ waveforms allocated for memory fading, $50$ waveforms for training, and $50$ waveforms for testing. 
(a) A segment of the input function, $f_{k}$, showing the first $10$ waveforms during the testing phase. 
(b) The corresponding segment of the actual input applied to the reservoir, $\widetilde{f}_{k}$, modified by the use of $4$ feedback channels.
(c)-(f) The corresponding readouts from a single-atom reservoir with $\omega_{1}=20$ and $g_{1}=30$. 
Parameters: $\omega_{c}=40$, $\kappa=10$, $\epsilon=20$, $\omega_{ss}=10$, $N_{ss}=16$.}
\label{Sine_Square_Input_Readouts}
\end{figure}

\begin{figure*}
\includegraphics[width=1.0\linewidth]{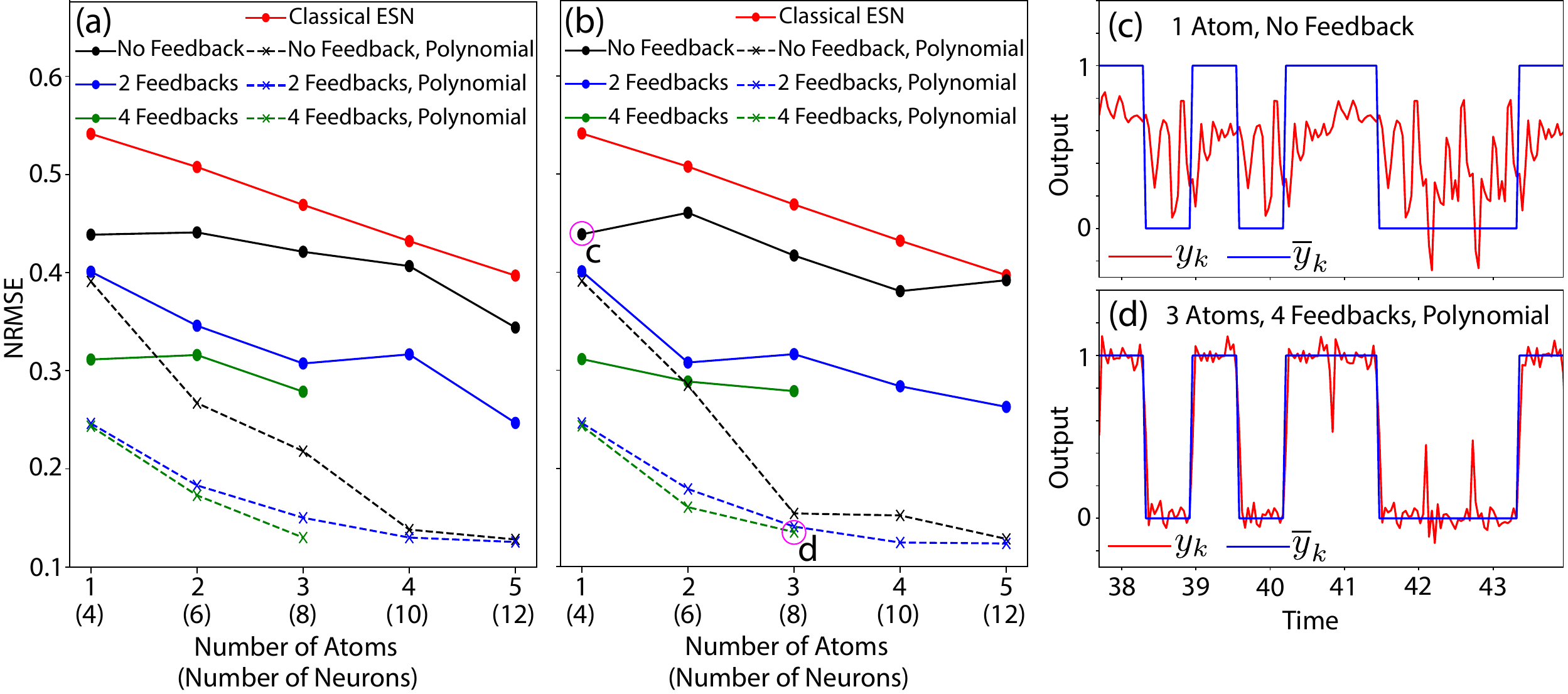}
\caption{Performance testing for the sine-square waveform classification task with various reservoir scales. 
(a)(b) Tested NRMSE plotted against the number of atoms (or neurons), using all available readout channels, $0$, $2$, or $4$ feedback channels, and regular linear regression (solid lines) or polynomial regression (dashed lines), with a comparison to CRC employing classical ESN (red solid line), where the parameters $\omega_{i}$ and $g_{i}$ align with those in Fig.~\ref{Mackey_Glass_Change_AtomNum}(a)(b). 
(c)(d) Actual (red) and target (blue) outputs from QRC utilizing one atom with no feedback with linear regression, and three atoms with $4$ feedbacks with polynomial regression, corresponding to the points marked by ``c'' and ``d'' in panel (b), respectively. 
Parameters: $\omega_{c}=40$, $\kappa=10$, $\epsilon=20$, $\omega_{ss}=10$, and $N_{ss}=16$.}
\label{Sine_Square_Change_AtomNum}
\end{figure*}

\subsection{Sine-square Waveform Classification Task}

The objective of the classification task is to determine whether each input data point belongs to a sine or a square waveform. The time-dependent input, $f_{k}$, comprises $110$ randomly generated sine and square waveforms. Among these, $10$ waveforms are allocated for memory fading, $50$ for training, and $50$ for testing. Each waveform is discretized into $N_{ss}$ points, leading to a time step of $\Delta t=2\pi/\left(N_{ss}\omega_{ss}\right)$, where $\omega_{ss}$ denotes the oscillation frequency of the input. Figure~\ref{Sine_Square_Input_Readouts}(a) depicts these discretized time points (red dots) across the first $10$ waveforms during the testing phase. The corresponding input modified by feedback is plotted in Fig.~\ref{Sine_Square_Input_Readouts}(b). The target output, $\bar{y}_{k}$, aimed at classifying the input signal, is set to $0$ if the input point belongs to a square waveform, and $1$ if it belongs to a sine waveform. Compared to the Mackey-Glass task, the sine-square task requires shorter-term memory, since the extremal points require memory to be distinguished--input values of $0$ and $2$ can correspond to either the sine or the square wave.

The sine-square waveform classification is a nonlinearity task that requires the reservoir to process a linearly inseparable input dataset containing abrupt changes. In a linear, closed quantum system, the smoothly evolving expectation values in readouts, $\left\langle c\right\rangle \propto\exp\left(-i\omega_{c}t\right)$ and $\left\langle \sigma_{i}\right\rangle \propto\exp\left(-i\omega_{i}t\right)$, are unable to capture these high-frequency, abrupt shifts. The introduction of nonlinearity, originating from the coherent laser driving and decay in the open quantum system, enhances the reservoir's capability to promptly respond to the abrupt changes in the input. This is evidenced by the distinct readout measurements in Fig.~\ref{Sine_Square_Input_Readouts}(c)-(f), which correspond directly to the sine and square input waveforms in Fig.~\ref{Sine_Square_Input_Readouts}(a). The various feature shapes observed in the readout measurements highlight the reservoir's capability to rapidly respond to the sudden shifts in input.

The performance enhancements resulted from the increased number of atoms, feedback mechanism, and polynomial regression, along with the comparison between QRC and CRC, are shown in Figs.~\ref{Sine_Square_Change_AtomNum}(a)(b). It is observed that the performance associated with polynomial regression, represented by the dashed lines, tends to saturate at $4$ atoms. This performance saturation suggests that the maximum performance achievable with the current sample size, associated with $N_{ss}=16$, has been attained. The enhanced performance is illustrated in Fig.~\ref{Sine_Square_Change_AtomNum}(d), where discrepancies between the actual and target outputs primarily occur at the first point in each waveform. These discrepancies characterize the short-time responses of the quantum reservoir to the abrupt shifts in the input waveform. The dependence of QRC performance on various samples sizes $N_{ss}$ and the influence of abrupt shifts in the input are detailed in the ``Supplementary Information''.

\section{Discussion}

We have introduced a paradigm for minimalistic quantum reservoirs that enhances quantum memory and nonlinear data processing capabilities. This approach incorporates a feedback mechanism that modifies the inputs based on observable readouts. Specifically, the inputs are encoded into the intensity of the cavity-driving laser, enabling a practical implementation of feedback by adjusting the laser's intensity. In a previous work, a Hamiltonian configuration strategy was proposed to enhance QRC performance by using feedback from the outputs to modify the reservoir dynamics \cite{Xia2023}. This approach is based on controlling internal parameters of the quantum reservoir, analogous to adjusting the atom-photon coupling strength $g_{i}$ and the detuning $\omega_{i}$ in our system. In contrast, our approach does not involve adjusting these internal parameters. Instead, the parameter $V_{n_f}$ is treated as an external parameter that lies outside the quantum reservoir hardware. The primary purpose of our external feedback is to provide an enhanced computational complexity by easily adjusting the weights $V_{n_f}$ through mostly software methods. This purpose is consistent with the original QRC paradigm, which does not require optimization of internal parameters.

To further augment the system, we employ polynomial regression to append quadratic combinations of observable expectations to the readouts, introducing an additional layer of complexity and capability. Similar nonlinear combinations of measured expectations have been leveraged in previous proposals based on qubit networks \cite{CNY20, chen2019learning}. However, in these qubit-based systems, incorporating second- and third-order polynomial terms into the readouts has not led to significant improvements in NRMSE \cite{chen2019learning}. In contrast, our photon-qubit coupled system benefits from the high-dimensional photon subspace, which enhances the reservoir's expressive capacity. This potential is effectively harnessed through polynomial regression.

These novel components significantly enhance the performance of QRC for small reservoirs containing just one to five atoms. To obtain the readouts in practice, we propose the use of continuous quantum measurements via homodyne detection of cavity quadratures and atomic spins. This approach allows for the simultaneous measurement of non-commuting observables \cite{Wei2008,Ruskov2010,Hacohen2016,Ochoa2018}, avoiding the need for tomography and making it more practical than the measurement of probability distributions in quantum basis states suggested in prior works \cite{Fujii2017,Ghosh2019,Dudas2023,Angelatos2021}.

Our proposed quantum optical reservoir offers convenient scalability compared to reservoirs built upon quantum networks like the Ising \cite{Fujii2017} and Fermi-Hubbard \cite{Ghosh2019} models. This advantage arises from the practical coupling between atoms and a single-mode cavity field in our system. Specifically, to scale up the reservoir, we simply need to add more atoms into the cavity. These atoms naturally couple with the cavity field, which in turn automatically couples them with the rest of the atoms in the reservoir.

Moreover, the number of quantum basis states in our reservoir scales exponentially with the number of atoms, following $2^{N_{\text{atom}}}$. 
This leads to faster growth compared to reservoirs that are not based on quantum networks. In systems such as a single Kerr nonlinear oscillator \cite{Govia2021} or two coupled linear oscillators \cite{Dudas2023}, the number of basis states increases linearly with the energy levels involved in the dynamics.
The exponential scaling of our reservoir has been crucial in demonstrating the advantages of QRC over CRC, as shown in Figs~\ref{Mackey_Glass_Change_AtomNum}(a)(b) and \ref{Sine_Square_Change_AtomNum}(a)(b). To compare with CRC using the same number of neurons, Fig.~\ref{Mackey_Glass_Change_AtomNum}(a) shows that QRC with 3 atoms significantly improves performance by reducing the NRMSE by $51\%$. For a 3-atom QRC, the introduction of a feedback mechanism with 4 feedback channels reduces the NRMSE by $42\%$, while polynomial regression alone contributes a $32\%$ reduction. When both the feedback mechanism and polynomial regression are applied together, the NRMSE reduction increases to $65\%$, as illustrated in Fig.~\ref{Mackey_Glass_Change_AtomNum}(a).

The cooperativity serves as a key metric for assessing the feasibility of the proposed QRC scheme on realistic experimental platforms. It is a dimensionless parameter defined as $g_{i}^2 / \kappa_{c} \kappa_{i}$, where $g_{i}$ denotes the atom-photon coupling strength, $\kappa_{c}$ is the cavity photon loss rate, and $\kappa_{i}$ is the atomic spontaneous emission rate. In our simulations, we use typical values of $g_{i} = 10$ and $\kappa_{c} = \kappa_{i} = 5$, resulting in a single-atom cooperativity of approximately $4$. Given that many experimental platforms operate with cooperativity values ranging from $1$ to $10$ \cite{Reitz2022}, our simulated system is within realistic experimental conditions.

\section{Methods}

\subsection{Continuous quantum measurement for observable readouts}

Continuous quantum measurements are simulated for both the cavity field and individual atoms. For the cavity field, homodyne detection of two orthogonal quadratures involves splitting the system's output beam into two using a beam-splitter, followed by homodyning each beam with the same local oscillator, differing by a phase shift of $\pi/2$ \cite{Wiseman2010}. Similarly, for atoms, homodyne detection is performed for spontaneous emissions \cite{Wiseman2001}. These measurements can be conducted concurrently \cite{Wei2008,Ruskov2010,Hacohen2016,Ochoa2018}. Consequently, the continuous measurement process is described by the stochastic master equation \cite{VPB79,VPB88,VPB91a,HC93,Wiseman2010,BvHJ07,Nurdin14}
\begin{align}
d\rho_{J} & =-i\left[H_{0}+H_{1}\left(t\right),\rho_{J}\right]dt\label{eq:sme}\\
 & +2\mathcal{D}\left[\sqrt{\kappa_{c}}c\right]\rho_{J}dt+2\underset{i}{\sum}\mathcal{D}\left[\sqrt{\kappa_{i}}\sigma_{i}\right]\rho_{J}dt\nonumber \\
 & +\left(dW_{Q}\mathcal{H}\left[\sqrt{\kappa_{c}}c\right]+dW_{P}\mathcal{H}\left[i\sqrt{\kappa_{c}}c\right]\right)\rho_{J}\nonumber \\
 & +\underset{i}{\sum}\left(dW_{x,i}\mathcal{H}\left[\sqrt{\kappa_{i}}\sigma_{i}\right]+dW_{y,i}\mathcal{H}\left[i\sqrt{\kappa_{i}}\sigma_{i}\right]\right)\rho_{J},\nonumber 
\end{align}
where the deterministic part is governed by the Lindblad superoperator $\mathcal{D}$ defined as 
\begin{equation}
\mathcal{D}\left[a\right]\rho_{J}=a\rho_{J} a^{\dagger}-\frac{1}{2}\left(a^{\dagger}a\rho_{J}+\rho_{J} a^{\dagger}a\right),\label{eq:Lindblad}
\end{equation}
and the stochastic part is determined by the superoperator $\mathcal{H}$ defined as
\begin{equation}
\mathcal{H}\left[a\right]\rho_{J}=a\rho_{J}+\rho_{J}a^{\dagger}-\left\langle a+a^{\dagger}\right\rangle _{J}\rho_{J}\label{eq:Stochastic}
\end{equation}
for any stochastic collapse operator $a$. The continuous quantum measurements of the observables $Q$, $P$, $\sigma_{x,i}$, and $\sigma_{y,i}$ are associated with the stochastic collapse operators $\sqrt{\kappa_{c}}c$, $i\sqrt{\kappa_{c}}c$, $\sqrt{\kappa_{i}}\sigma_{i}$, and $i\sqrt{\kappa_{i}}\sigma_{i}$, respectively \cite{Wiseman2010,BvHJ07,Nurdin14}. The randomness of the measurement records is taken into account by the Wiener increments, $dW_{Q\left(P\right)}$ and $dW_{x\left(y\right),i}$, each of which selects a random number from a Gaussian probability distribution with a width of $dt$. The efficiencies of various detection channels are incorporated into the Wiener increments. Each measurement detects the continuous currents in cavity and atom channels with noises, with the measurement records given by $\left\langle Q\left(P\right)\right\rangle _{J}+dW_{Q\left(P\right)}/dt$ and $\left\langle \sigma_{x\left(y\right),i}\right\rangle _{J}+dW_{x\left(y\right),i}/dt$, where the expectation values are computed using $\rho_{J}$ from Eq.~(\ref{eq:sme}). The ``smesolve'' method in the Python QuTiP library is employed to solve the stochastic master equation (\ref{eq:sme}).

As the number of measurements approaches infinity, the impacts of the measurement back-actions, $dW_{Q\left(P\right)}$ and $dW_{x\left(y\right),i}$, are averaged out. This idealization is described by the deterministic master equation
\begin{equation}
\frac{d\rho}{dt}=-i\left[H_{0}+H_{1}\left(t\right),\rho\right]+2\mathcal{D}\left[\sqrt{\kappa_{c}}c\right]\rho+2\underset{i}{\sum}\mathcal{D}\left[\sqrt{\kappa_{i}}\sigma_{i}\right]\rho,\label{eq:me}
\end{equation}
accompanied by the averaged measurement records $\left\langle Q\left(P\right)\right\rangle $ and $\left\langle \sigma_{x\left(y\right),i}\right\rangle $, where the expectation values are computed using $\rho$ from Eq.~(\ref{eq:me}). To maintain a constant total decay rate $\kappa$ as $N_{atom}$ increases, the decay rate of the cavity or each atom is assumed to be $\kappa_{c\left(i\right)}=\kappa/\left(2N_{atom}+2\right)$. A higher $\kappa$ reduces the uncertainty in the readout measurement, commonly referred to a strong measurement, while a lower $\kappa$ corresponds to a weak measurement \cite{Fuchs2001}. The influence of the magnitude of $\kappa$ on QRC performance is demonstrated in Fig.~\ref{Mackey_Glass_Change_Delay_Kappa}. The ``mesolve'' method in the 
Python QuTiP library is employed to solve the deterministic master equation (\ref{eq:me}).

\begin{figure}
\includegraphics[width=1.0\linewidth]{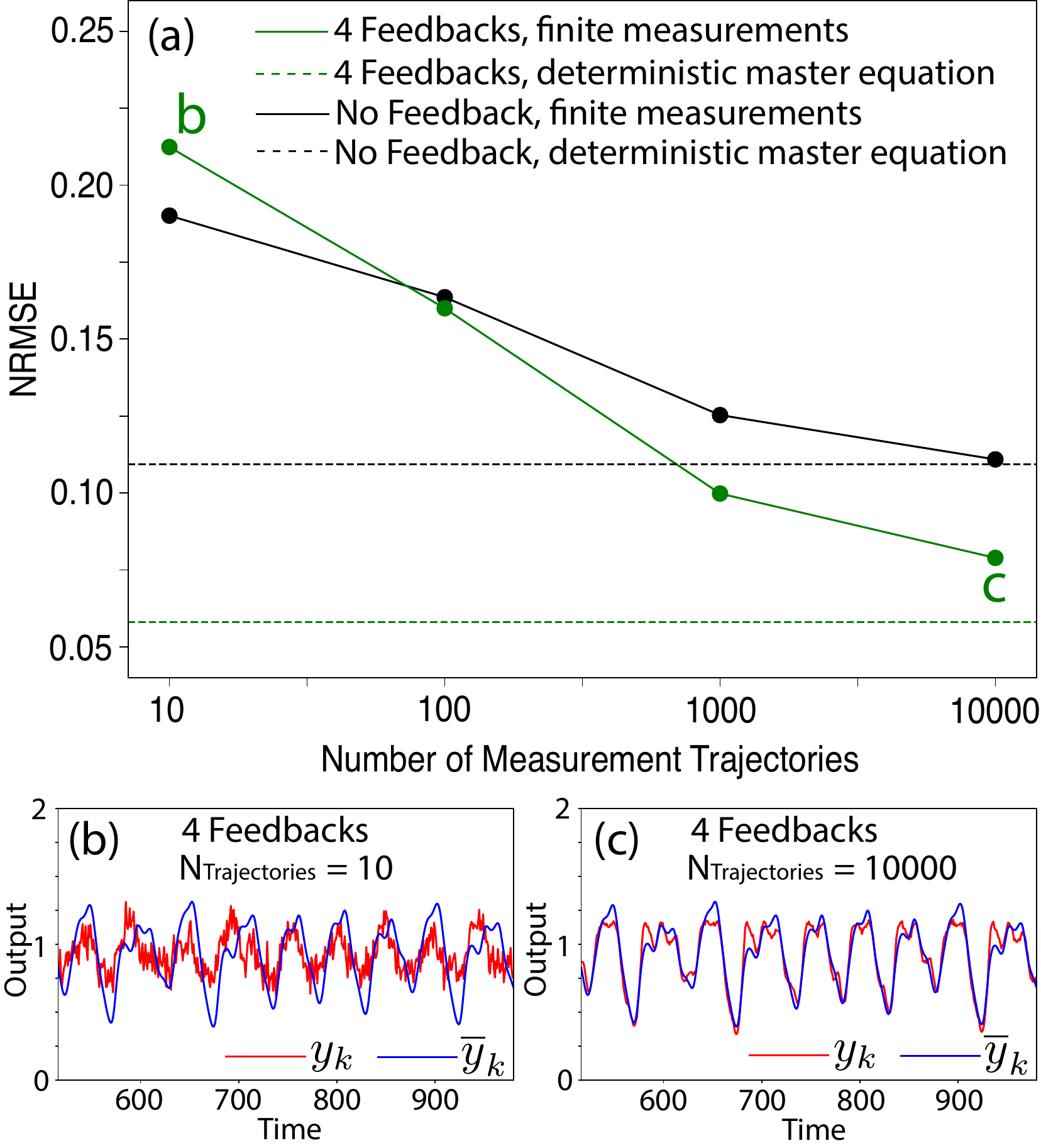}
\caption{Testing results for the Mackey-Glass task within the framework of continuous quantum measurement, where each readout is obtained by averaging measurement records over multiple measurement trajectories, with each measurement trajectory running through the whole time domain. (a) NRMSE as a function of the number of measurement trajectories for a single-atom QRC with $g_{1}=30$ and $\omega_{1}=20$. Solid lines: averaged results from a finite number of trajectories simulated with stochastic master equation. Dashed lines: asymptotic results simulated with deterministic master equation. Green lines: $4$ feedbacks. Black lines: No feedback. (b)(c) The actual (red) and target (blue) outputs for $10$ and $10000$ trajectories with $4$ feedbacks, corresponding to the points labeled ``b'' and ``c'' in panel (a), respectively. Parameters: $Delay=20$, $\kappa=10$, $\omega_{c}=40$, $\epsilon=20$.}
\label{Mackey_Glass_Trajectories}
\end{figure}
 
To eliminate the need for many copies of quantum systems and many repetitions of measurement to estimate observable expectations at each time step, we propose an online protocol based on continuous quantum measurement. This method requires only a single quantum system and generates multiple quantum trajectories, each evolving over the entire physical time domain, thereby avoiding destructive measurements. These trajectories are implemented sequentially on the same quantum system, with a memory-fading time interval introduced between each pair of successive trajectories. The fading memory feature of our QRC scheme eliminates the need to reset the initial state before the start of each trajectory.  

For the QRC scheme without feedback, a continuous measurement trajectory is simulated by evolving the stochastic master equation in Eq.~(\ref{eq:sme}) over the entire time domain from $t = 0$ to $t = L \Delta t$, where $L$ is the maximum value of $k$ in the discretized time $t = k \Delta t$. The test phase of QRC is finalized at $t = L \Delta t$. The required number of experimental repetitions is equal to the number of measurement trajectories, denoted as $N_{Trajectories}$. To simulate the stochastic nature of measurement noise, the stochastic master equation in Eq.~(\ref{eq:sme}) is numerically evolved multiple times, each with independent realizations of the random Wiener increments $dW_{Q\left(P\right)}$ and $dW_{x\left(y\right),i}$. Each full evolution corresponds to a single measurement trajectory. The observable readouts, $x_{kn}$, are then obtained by averaging the measurement records across all such trajectories. The effect of using a finite number of measurement trajectories is illustrated by the black solid line in Fig.~\ref{Mackey_Glass_Trajectories}(a) for a single-atom QRC without feedback. The idealized case, based on observables averaged over an infinite number of trajectories, is simulated using the deterministic master equation (\ref{eq:me}), with the result plotted by the flat black dashed line in Fig.~\ref{Mackey_Glass_Trajectories}(a). As shown in Fig.~\ref{Mackey_Glass_Trajectories}(a), increasing the number of trajectories leads to improved QRC performance, with the result for $N_{Trajectories} = 10000$ approaching the limit obtained from an infinite number of trajectories. In the ``Results'' section, we use the deterministic master equation method to characterize QRC performance.

For the QRC scheme enhanced with feedback, our continuous measurement approach also allows each measurement trajectory to span the entire time domain from $t = 0$ to $t = L \Delta t$ without destroying the quantum reservoir. This eliminates the need for a large number of measurement repetitions to estimate observable expectations at each time step. In the feedback scheme, the required number of experimental repetitions remains equal to the number of measurement trajectories. For each trajectory, the system evolves from $t_k$ to $t_{k+1}$ according to the stochastic master equation (\ref{eq:sme}), during which the modified input and density operator are updated from $\widetilde{f}_{k}$ and $\rho_{J}\left(t_{k}\right)$ to $\widetilde{f}_{k+1}$ and $\rho_{J}\left(t_{k+1}\right)$, respectively. At the time $t_{k+1}$, the evolution is uninterrupted as we only take one non-destructive measurement shot from the current trajectory. To ensure the stability of the feedback mechanism, the modified input $\widetilde{f}_{k}$ for the $M$-th trajectory is determined by
\begin{equation}
\widetilde{f}_{k}=f_{k}+\underset{n_{f}}{\sum}\overline{x}_{kn_{f}}^{M}V_{n_{f}},\label{eq:fmeasure}
\end{equation}
where $\overline{x}_{kn_{f}}^{M}$ is the averaged readouts from all previous measurement trajectories indexed from $1$ to $(M-1)$, which is given by
\begin{equation}
\overline{x}_{kn_{f}}^{M}=\frac{1}{M-1}\mathop{\sum_{m=1}^{M-1}}x_{kn_{f}}^{m},\label{eq:xmeasure}
\end{equation}
with $m$ denoting the index of a previous trajectory and $x_{kn_{f}}^{m}$ the readout from the $m$-th trajectory. Note that the readouts from the current ($M$-th) trajectory, denoted as $x_{kn_{f}}^{M}$, are determined by the solution $\rho_{J}\left(t_{k}\right)$ from the stochastic master equation of the current trajectory. The idealized case for feedback QRC is simulated using the deterministic master equation (\ref{eq:me}), using the feedback modified input governed by Eq.~(\ref{eq:ftilde}). In the ``Results'' section, we use the deterministic master equation to evaluate the performance of QRC enhanced with feedback.

The performance of feedback QRC with a finite number of measurement trajectories is illustrated by the green solid line in Fig.~\ref{Mackey_Glass_Trajectories}(a). The averaged measurement result asymptotically approaches the NRMSE obtained using the deterministic master equation. Notably, measurement noise has a more pronounced effect on the feedback scheme compared to the scheme without feedback, as evidenced by the larger gap between the green solid and dashed lines than that between the black solid and dashed lines in Fig.~\ref{Mackey_Glass_Trajectories}(a). This observation is consistent with the fact that measurement noise affects not only the readouts from the quantum reservoir but also the inputs modified via feedback based on those readouts. Strikingly, as shown in Fig.~\ref{Mackey_Glass_Trajectories}(a), even in the presence of measurement noise, the feedback-enhanced QRC outperforms the idealized QRC without feedback when the number of trajectories exceeds $1000$.

\subsection{Training quantum reservoir computing with feedback mechanism}

\label{sec:NRMSEtraining}

The objective of training is to optimize the parameters $V_{n_{f}}$ and $W_{n}$ in order to achieve the actual output $y_{k}$ such that the NRMSE defined in Eq.~(\ref{eq:NRMSE}) is minimized.
 
To train the parameters $W_{n}$, a pseudoinverse method is sufficient in the regression analysis, since ${\rm NRMSE^{2}}$, being a quadratic function of $W_{n}$, only has a global minimum but no local minima. The readouts, $x_{kn}$, along with the constant bias term $x_{k0}=1$, are arranged in an $L\times\left(N_{readouts}+1\right)$ matrix ${\bf X}$. The target output, $\bar{y}_{k}$, is organized in an $L\times1$ column vector ${\bf {\bf \bar{Y}}}$. The weights, $W_{n}$, are arranged in a $\left(N_{readouts}+1\right)\times1$ column vector ${\bf W}$. The optimized weight ${\bf {\bf W}}$ that minimizes NRMSE is therefore determined by \cite{Dion2018,Govia2021}
\begin{equation}
{\bf W}={\bf X}^{+}{\bf \bar{Y}},\label{eq:W}
\end{equation}
where the Moore-Penrose pseudoinverse 
\begin{equation}
{\bf X}^{+}=\left({\bf X^{{\rm T}}}{\bf X}+\delta{\bf I}\right)^{-1}{\bf X^{{\rm T}}},\label{eq:MoorePenrose}
\end{equation}
with ${\bf I}$ being the identity square matrix, and $\delta=10^{-10}$ is a ridge-regression parameter used to prevent overfitting. 

After the optimization regarding $W_{n}$ for any given $V_{n_{f}}$, NRMSE can be regarded as a function of $V_{n_{f}}$. To optimize the parameter $V_{n_{f}}$, we adopt three global optimization methods in the SciPy library: differential evolution, brute force, and brute force plus Nelder Mead. In any of these three global optimization methods with any number of readouts and feedbacks, the searching boundaries are set to $V_{n_{f}}\in\left[-3,3\right]$ for each feedback channel to ensure a fair comparison between various cases. In the differential evolution method, maximum number of iterations is set to $1000$ and the lowest NRMSE is selected from three batch results due to the stochastic nature of this method. In the brute force method, the space of $V_{n_{f}}$ is sliced with the step size $0.5$, and the site that leads to the lowest NRMSE is selected as the optimized $V_{n_{f}}$. In the brute force plus Nelder Mead method, the local optimization method Nelder Mead is additional applied to further search for a local minimum around the global minimum resulted from the basic brute force. The three global optimization methods lead to similar results shown in Fig.~\ref{Mackey_Glass_Change_AtomNum}(h).

\section*{Acknowledgements}
We gratefully acknowledge the startup funding provided by Wyant College of Optical Sciences, University of Arizona, which supported the initial development of this research project.

\section*{Author Contributions}
C.Z. and D.S. conceived the initial quantum reservoir computing system concept. C.Z, H.N., and P.E. contributed to the detailed operational design of the system. C.Z. conducted all numerical simulations. All authors contributed equally to the manuscript preparation.

\bibliography{References_Practical_Scalable_QRC}

\end{document}


\title{Supplementary Information for Minimalistic and Scalable Quantum Reservoir Computing with Feedback}

\author[1]{\fnm{Chuanzhou} \sur{Zhu}}
\author[1]{\fnm{Peter J.} \sur{Ehlers}}
\author[2]{\fnm{Hendra I.} \sur{Nurdin}}
\author[1]{\fnm{Daniel} \sur{Soh}}
\affil[1]{\orgdiv{Wyant College of Optical Sciences}, \orgname{University of Arizona}, \orgaddress{\city{Tuscon}, \state{AZ}, \country{US}}}
\affil[2]{\orgdiv{School of Electrical Engineering and Telecommunications}, \orgname{University of New South Wales}, \orgaddress{\city{Sydney}, \country{Australia}}}

\maketitle

\section{Influence of Feedback Parameters on QRC Performance}

\begin{figure}
\centering
\includegraphics[width=1.0\linewidth]{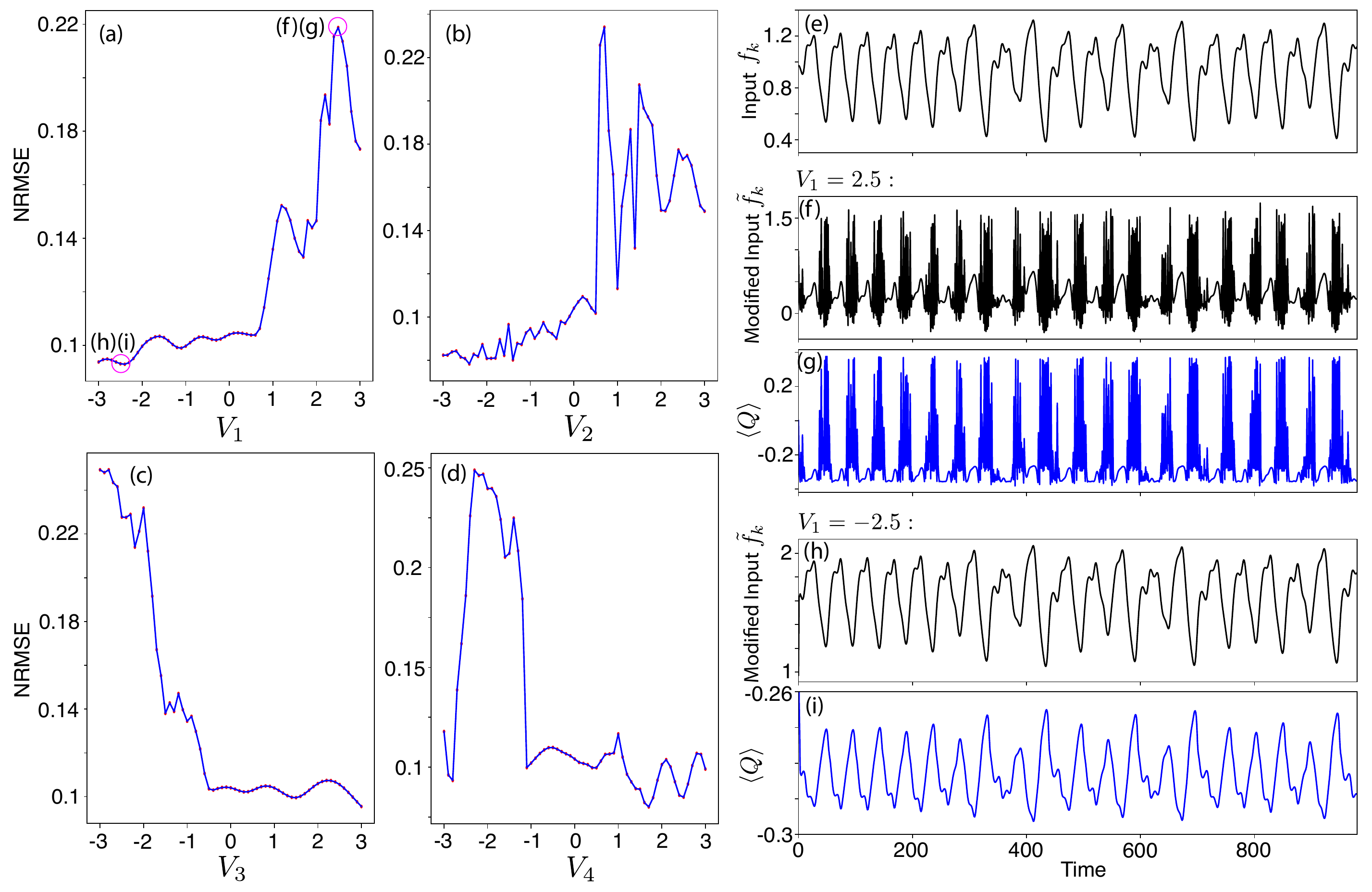}
\caption{Influence of various feedback parameters $V_{n_{f}}$ for the Mackey-Glass task utilizing a single-atom reservoir with $\omega_{1}=20$ and $g_{1}=30$. (a)-(d) NRMSE as a function of the single feedback parameter, $V_{1}$, $V_{2}$, $V_{3}$ or $V_{4}$, with only one feedback channel, $x_{k1}$, $x_{k2}$, $x_{k3}$, or $x_{k4}$, turned on and the rest three feedback channels turned off, where the parameters $W_{n}$ have been trained for each specified $V_{n_{f}}$ using regular linear regression and all the four available readouts. (e) The Mackey-Glass input $f_{k}$. (f)(h) The actual input applied to the reservoir, $\widetilde{f}_{k}$, modified by the feedback mechanism using $V_{1} = 2.5$ and $V_{1} = -2.5$, respectively. (g)(i) The corresponding readout $x_{k1}=\left\langle Q\right\rangle _{k}$ for $V_{1} = 2.5$ and $V_{1} = -2.5$, respectively. The feedback parameters for (f)(g) and (h)(i) are marked by the pink circles in panel (a). Parameters: $Delay=20$, $\kappa=10$, $\omega_{c}=40$, and $\epsilon=20$.}
\label{Supplementary_One_Feedback}
\end{figure}

As elaborated in the ``Methods" section, for the purposes of training, the NRMSE can be regarded as a function of the feedback parameters $V_{n_{f}}$ after the optimized $W_{n}$ have been obtained for each value of $V_{n_{f}}$. For a single-atom QRC, 4 readouts can be obtained from the measurement of expectation values as $x_{k1}=\left\langle Q\right\rangle _{k}$, $x_{k2}=\left\langle P\right\rangle _{k}$, $x_{k3}=\left\langle \sigma_{x,1}\right\rangle _{k}$, $x_{k4}=\left\langle \sigma_{y,1}\right\rangle _{k}$. The maximal utilization of these 4 readouts as feedback channels lead the NRMSE to be a function of the feedback coefficients $V_{1}$, $V_{2}$, $V_{3}$ and $V_{4}$. The global minimum in this 4-dimensional space gives a tested NRMSE right below $0.06$, according to leftmost green dot in Fig.~3(a) in the main text, with the modified input and reservoir dynamics shown in Fig.~2 in the main text. The cross section of the NRMSE along each of the 4 axes is illustrated here in Figs.~\ref{Supplementary_One_Feedback}(a)-(d), where the NRMSE is significantly higher in some parameter regions. In these higher-NRMSE regions, as plotted in Figs.~\ref{Supplementary_One_Feedback}(f)(g), we observe high-frequency oscillations in the modified input and readouts, which can be regarded as a limit cycle like behavior. In the lower-NRMSE regions, as depicted in Figs.~\ref{Supplementary_One_Feedback}(h)(i), the high-frequency oscillations disappear and the reservoir dynamics are more similar to the optimized result shown in Fig.~2 in the main text.

\section{Classical Reservoir Computing on Echo State Networks}

\begin{figure}
\centering
\includegraphics[width=0.8\linewidth]{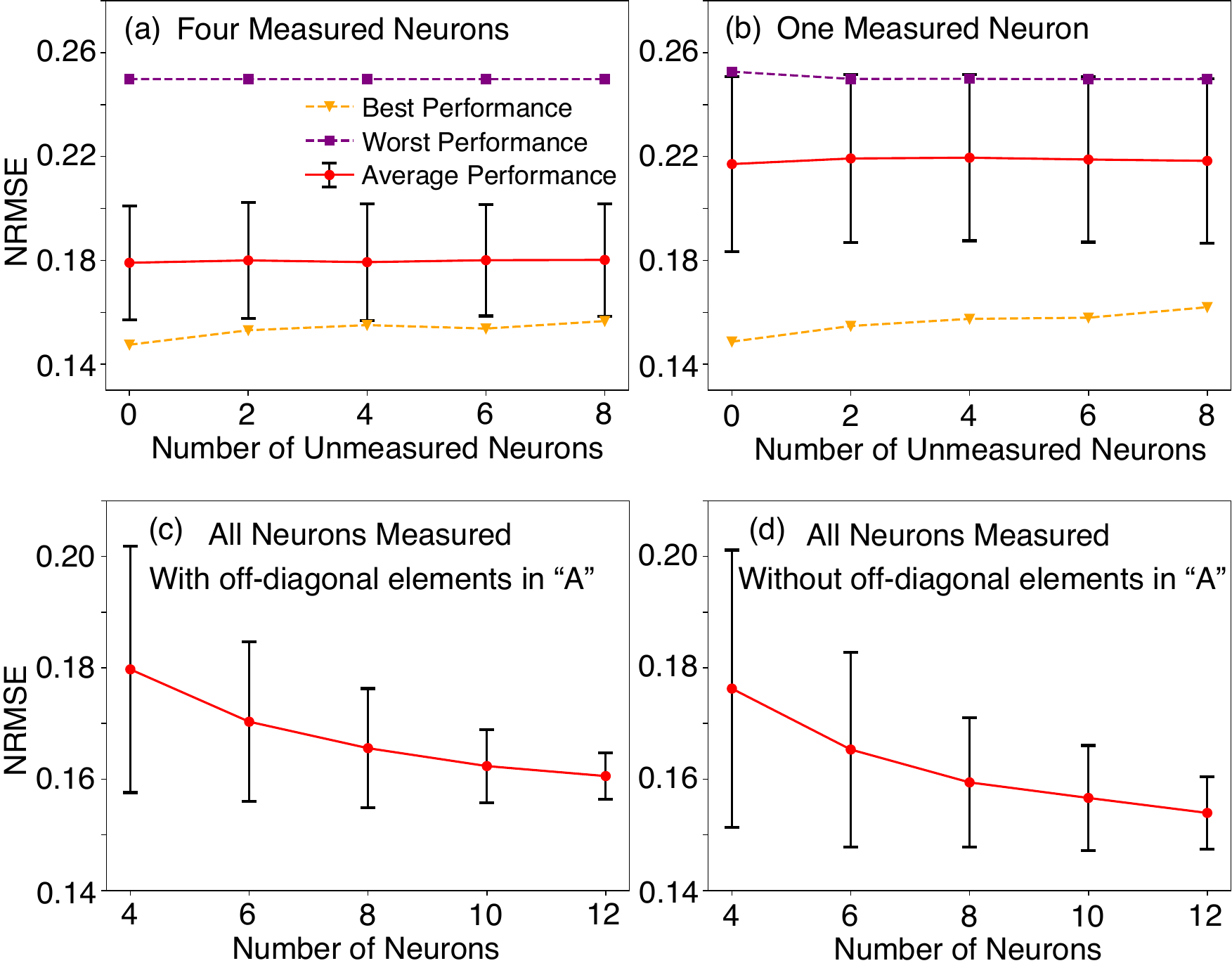}
\caption{Tested NRMSE for the Mackey-Glass task using classical ESNs by averaging $1000$ random trajectories ($1000$ random $A$ and $B$ matrices), with the error bars and the best and worst performances plotted in account of the randomness. (a) The number of measured neurons used for training and testing is fixed at $4$, while the number of unmeasured neurons increases from $0$ to $8$, with the corresponding dimensions of $A$ and $B$ matrices increasing from $4$ to $12$. (b) The number of measured neurons is fixed at $1$. (c)(d) All neurons are measured for training and testing, with off-diagonal elements in the $A$ matrix turned on in (c) and turned off in (d). The red lines in panel (a) and (c) correspond to the red lines for classical ESN in Figs.~3(a)(b) and Fig.~3(c) in the main text, respectively.}
\label{Supplementary_CRC}
\end{figure}

\begin{figure}
\centering
\includegraphics[width=0.6\linewidth]{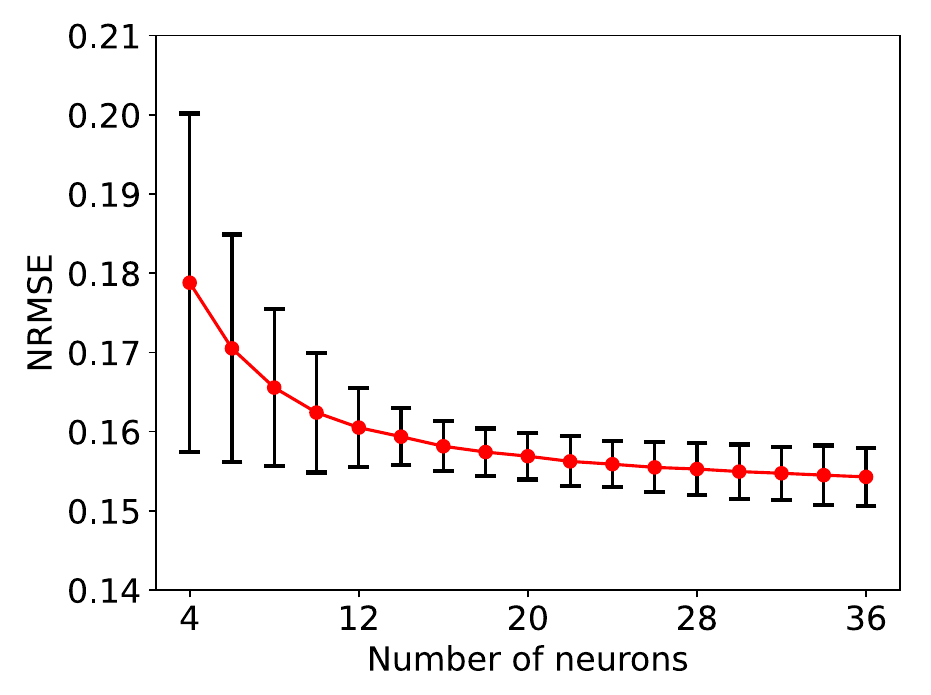}
\caption{Tested NRMSE for the Mackey-Glass task using classical ESNs over a broader range of the number of neurons, by averaging $1000$ random trajectories ($1000$ random $A$ and $B$ matrices), with the error bars plotted in account of the randomness. All other conditions are identical to those used in Fig.~\ref{Supplementary_CRC}(c) and the red solid line in Figs.~3(a)(b) of the main text.}
\label{Supplementary_CRC_larger_size_AB}
\end{figure}

For the classical reservoir computing (CRC) results, we use echo state networks (ESNs) described by 
\begin{align}
    {\bf x}_{k+1} &= {\rm ReLU}({\bf A x}_k + {\bf B} u_k), \label{eq:ESN} \\
    y_k &= {\bf W^{\rm T} x}_k + C,
\end{align}
where $x_k$ is the vector-valued readout of the ESN at time step $k$ with length $N_{readouts}$, $u_k$ is the input, ${\bf W}$ and $C$ are fitting parameters obtained through optimization, and $y_k$ is the output at $k$. The activation function we use is the rectified linear unit ${\rm ReLU}(z)=\max(0,z)$ for a scalar $z$, and for vector arguments ${\bf x}$ the function is applied to each element of the vector so that $({\rm ReLU}({\bf x}))_i={\rm ReLU}(x_i)=\max(0,x_i)$. An individual ESN is defined by the choice of the matrix ${\bf A}$ and the vector ${\bf B}$.

To compare CRC and QRC performance, we get an average over $1000$ ESNs by choosing ${\bf A}$ and ${\bf B}$ randomly, where each element in ${\bf A}$ and ${\bf B}$ is randomly chosen between $-1$ and $1$. Additionally, we rescale ${\bf A}$ so that its largest singular value is less than one, which ensures convergence of the ESN. The number of neurons $N_{neuron}$ in an ESN corresponds to both the vector length of ${\bf x}_k$ and the dimensions of ${\bf A}$ and ${\bf B}$. Training of ${\bf W}$ and $C$ is performed using the same linear regression technique as the one introduced in "Methods" for training $W_{n}$ in QRC. 

In the results of Figs.~3(a)(b) and Figs.~6(a)(b) in the main text, all neurons in ESNs (and therefore all elements in ${\bf x}_k$) are measured for the training and testing purposes, meaning that the dimension of ${\bf W}$ is equal to $N_{neuron}$. Performance enhancement is observed as $N_{neuron}$ increases from $4$ to $12$. This result for the Mackey-Glass task is also reproduced in Fig.~\ref{Supplementary_CRC}(c), with the error bars plotted in account of the random ${\bf A}$ and ${\bf B}$ matrices.

In the result of Figs.~3(c) in the main text, the number of measured neurons is fixed at $4$, and therefore only the first $4$ elements in the state vector ${\bf x}_k$ are used for training and testing, meaning that the dimension of ${\bf W}$ is $4$. In this case, as the number of unmeasured neurons increases from $0$ to $8$, the corresponding dimensions of $A$ and $B$ matrices increase from $4$ to $12$. The result suggests that the correlations between the measured neurons and the newly added unmeasured neurons do not contribute to the performance enhancement. This result is further confirmed by Fig.~\ref{Supplementary_CRC}(a)(b) showing the best and worst performances and the error bars with the number of measured neurons being $4$ and $1$, respectively. According to Eq. (\ref{eq:ESN}), the correlations among neurons in an ESN are characterized by the off-diagonal elements in the $A$ matrix. In Fig.~\ref{Supplementary_CRC}(d), these correlations are turned off by setting all off-diagonal elements in $A$ to $0$. The comparison between Fig.~\ref{Supplementary_CRC}(c) (with off-diagonal elements in $A$) and Fig.~\ref{Supplementary_CRC}(d) (without off-diagonal elements in $A$) supports that the correlations among neurons in ESNs provide no help to enhance CRC performance. 

For classical ESNs with unmeasured neurons from 0 to 8, we also looked into possible dependencies of the performance on the spectral radius of $A$, the strength of the input, and the condition number of the Gram matrix ${\bf X^{{\rm T}}}{\bf X}$ used in training (see the ``Methods" section). We found no clear correlation between the average NRMSE as a function of unmeasured neurons and any of these parameters. Since there is no practical purpose in training an ESN using some but not all of the readouts, we have decided not to further pursue the underlying cause of the unchanging NRMSE as a function of unmeasured nodes.

To compare the performance of QRC with that of a larger classical ESN, we extend our simulation to include more classical nodes into CRC. This is achieved by increasing the dimensions of the random matrices $A$ and $B$ as well as the number of classical readouts. The result is illustrated in Fig.~\ref{Supplementary_CRC_larger_size_AB}, showing that the optimal performance attained by CRC ($\rm{NRMSE} \geq 0.15$) still falls behind the performance of a single-atom QRC without the enhancement from the feedback mechanism or polynomial regression ($\rm{NRMSE} \approx 0.11$, according to Fig.~3(a) of the main text).

\section{Sample Size in Sine-Square Waveform Classification Task}

\begin{figure}
\centering
\includegraphics[width=0.8\linewidth]{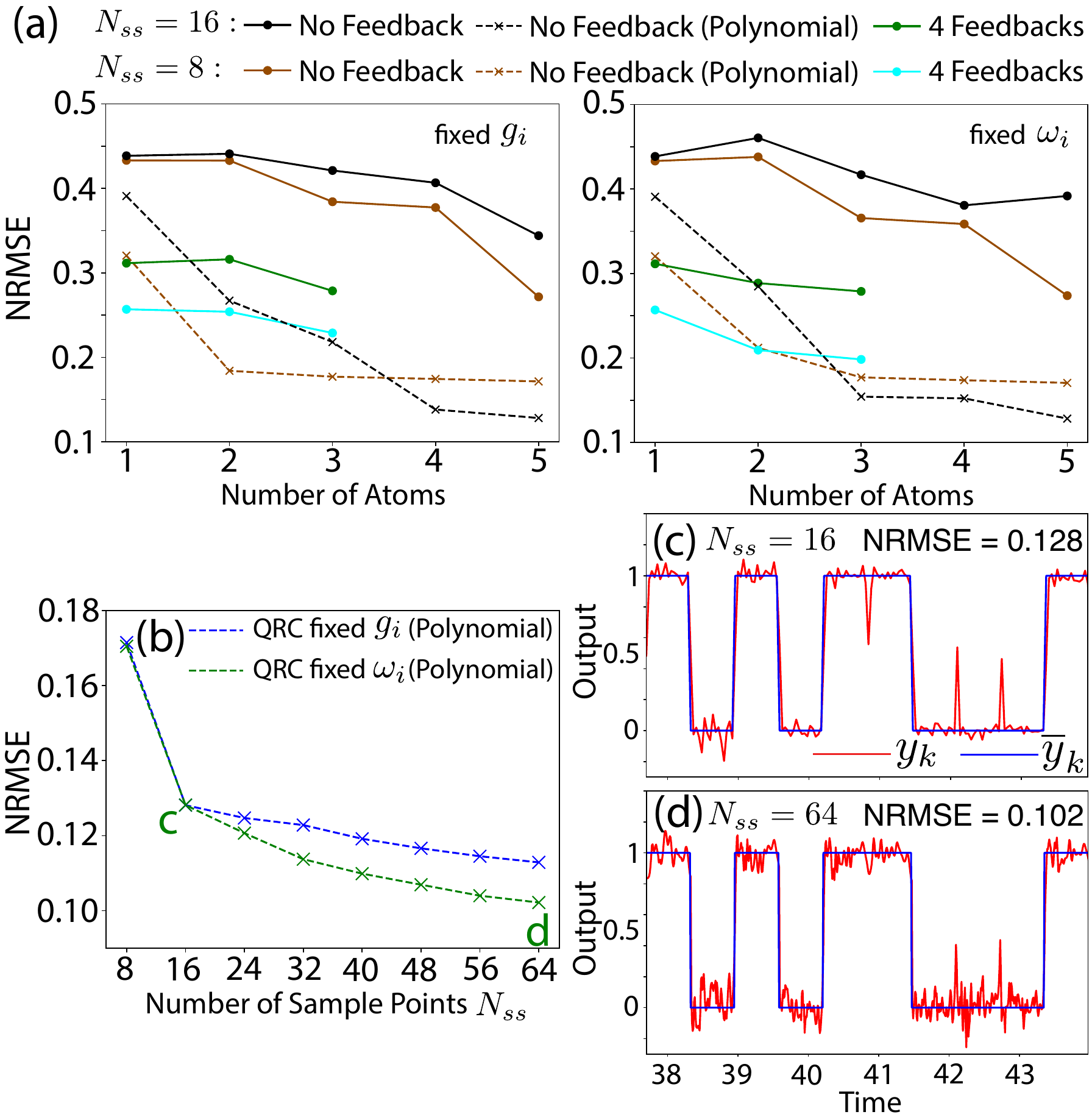}
\caption{Tested performance for the sine-square waveform classification task with various numbers of sample points, $N_{ss}$, within each period of the sine or square waveform input. 
(a) For the sample sizes $N_{ss}=16$ and $N_{ss}=8$, NRMSE plotted against the number of atoms, using all available readouts, no feedback (black and brown lines) or $4$ feedbacks (green and cyan lines), and regular linear regression (solid lines) or polynomial regression (dashed lines), with $g_{i}$ fixed in (a) and $\omega_{i}$ fixed in (b). Note that the results for $N_{ss}=16$ correspond to the black and green lines in Fig. 6(a)(b) in the main text. 
(b) NRMSE as a function of $N_{ss}$ for two sets of five-atom QRCs with $g_{i}=30$ and $\omega_{i}=[0,10,20,30,40]$ (blue) and with $\omega_{i}=20$ and $g_{i}=[10,20,30,40,50]$ (green), utilizing polynomial regression and no feedback. 
(c)(d) The actual (red) and target (blue) outputs with $N_{ss}=16$ and $N_{ss}=64$, corresponding to the "c" and "d" markers in panel (b), respectively.
Parameters: $\omega_{c}=40$, $\kappa=10$, $\epsilon=20$, and $\omega_{ss}=10$.}
\label{Supplementary_Sine_Square_Sample_Size}
\end{figure}

The sample size is another critical factor that influences QRC performance. In the sine-square waveform classification task, the sample size is quantified as the number of discretized time steps within each period of the sine or square waveform, denoted as $N_{ss}$. The comparison between the results for two difference sample sizes, $N_{ss}=16$ and $N_{ss}=8$, are given in Fig.~\ref{Supplementary_Sine_Square_Sample_Size}(a), where QRC schemes with regular linear regression, polynomial regression, and 4 feedbacks are considered as we increase the number of atoms by fixing $g_{i}$ and $\omega_{i}$. Generally, QRC performs better for $N_{ss}=8$, since a triangle-like waveform, compared to a sine-like waveform, is expected to be more different to a square waveform in its overall shape. However, the performance enhancement for $N_{ss}=8$ associated with polynomial regression (brown dashed lines) tends to stop when $N_{atom} \geq 3$. This performance limitation is related to the response of a quantum reservoir to abrupt changes in the input. In this specific region, the tested performance versus $N_{ss}$ for the five atom, polynomial regression case is plotted in Fig.~\ref{Supplementary_Sine_Square_Sample_Size}(b), and the comparisons between actual and target outputs for $N_{ss}=16$ and $N_{ss}=64$ are shown in Figs.~\ref{Supplementary_Sine_Square_Sample_Size}(c)(d). The result indicates that a larger $N_{ss}$, reducing the abruptness of the shift between two contiguous waveforms, requires a shorter reservoir's response time (which is characterized by the narrower pine peak in the actual output), and leads to better QRC performance.